\documentclass[11pt,a4paper]{article}

\usepackage{jcappub}
\usepackage{slashed}

\usepackage{multirow}
\usepackage{amssymb}
\usepackage{ulem}
\usepackage{cancel}
\usepackage{verbatim}
\usepackage{caption}
\usepackage{subcaption}
\usepackage{amsmath}
\usepackage{color}
\usepackage[utf8]{inputenc}
\usepackage{enumitem}

\definecolor{darkred}{rgb}{0.5,0,0}
\definecolor{darkgreen}{rgb}{0,0.5,0}
\definecolor{darkblue}{rgb}{0,0,0.5}

\usepackage{hyperref}
\hypersetup{ colorlinks,
linkcolor=darkblue,
filecolor=darkgreen,
urlcolor=darkred,
citecolor=darkblue }

\newcommand{\inspire}[1]{\href{http://inspirehep.net/search?p=find+J+#1}
 {{\color{black}[{\color{blue} {\small in}SPIRE}]}}}
\newcommand{\book}[1]{\href{http://inspirehep.net/search?p=#1}
 {{\color{black}[{\color{blue} {\small in}SPIRE}]}}}

\newcommand{\inspired}[1]{\href{http://inspirehep.net/search?p=#1}
 {{\color{black}[{\color{blue} {\small in}SPIRE}]}}}

\begin{document}

\title{Reionization and dark matter decay}

\date{\today}

\author{Isabel M.~Oldengott,}
\emailAdd{ioldengott@physik.uni-bielefeld.de}

\author{Daniel Boriero}
\emailAdd{boriero@physik.uni-bielefeld.de}

\author{and Dominik J. Schwarz}
\emailAdd{dschwarz@physik.uni-bielefeld.de}

\affiliation{Fakult\"at f\"ur Physik, Bielefeld University, D--33501 Bielefeld, Germany}

\abstract{Cosmic reionization and dark matter decay can impact observations of the cosmic microwave sky 
in a similar way. A simultaneous study of both effects is required to constrain unstable dark matter from 
cosmic microwave background observations. We compare two reionization models with and without dark 
matter decay. We find that a reionization model that fits also data from quasars and star forming 
galaxies results in tighter constraints on the reionization optical depth $\tau_{\text{reio}}$, but weaker 
constraints on the spectral index $n_{\text{s}}$ than the conventional parametrization. 
We use the Planck 2015 data to constrain the effective decay rate of dark matter to 
$\Gamma_{\rm eff} <  2.9 \times 10^{-25}/$s at $95$\% C.L. This limit is robust and model independent.
It holds for any type of decaying dark matter and it depends only weakly on the chosen 
parametrization of astrophysical reionization. For light dark matter particles that decay exclusively into 
electromagnetic components this implies a limit of $\Gamma <  5.3 \times 10^{-26}/$s at 
$95$\% C.L. Specifying the decay channels, we apply our result to 
the case of keV-mass sterile neutrinos as dark matter candidates and obtain constraints on their mixing 
angle and mass, which are comparable to the ones from the diffuse X-ray background.}

\maketitle   

\flushbottom
%%%%%%%%%%%%%%%%%%%%%%%%%%%%%%%%%%%%%%%%%%%%%%%%%%%%%%%%%%

\section{Introduction}

The universe is highly ionized today, although it was filled by cold 
gas of hydrogen and helium after the decoupling of photons. Evidence for reionization stems from 
observations of the spectra of distant quasars that show only a few absorption lines bluewards of 
the Lyman-$\alpha$ line in the quasar's rest frame. These absorption lines form the so-called 
Lyman-$\alpha$ forest can be understood as the fingerprints of neutral hydrogen clouds in the 
intergalactic medium. Already a very small amount of neutral hydrogen (fraction of $\sim 10^{-3}$) 
is sufficient to entirely suppress the quasar spectra. Hence, the fact that we still observe 
\textit{any} flux of quasars with $z \lesssim 6$ bluewards of the Lyman alpha line implies that the 
amount of neutral hydrogen must be very low and consequently the universe is ionized at $z\lesssim 6$. 

The observation of the Gunn-Peterson trough \cite{1965ApJ...142.1633G} (the absence of quasar radiation bluewards of the 
Lyman-$\alpha$ line) for the highest redshift quasars \cite{Becker:2001ee} furthermore indicates 
that the transition from a neutral to an ionized state happened at $z \sim 6$. 

A direct observation of the epoch of reionization could be obtained 
by means of the 21 cm transition of hydrogen \cite{2013ExA....36..235M,Pober:2013jna}. 
Instruments like the Low Frequency Array (LOFAR) \cite{Patil:2014dpa}, the Precision Array for Probing the Epoch of Reionization (PAPER) \cite{Ali:2015uua}
or the Murchison Widefield Array (MWA) \cite{PAS:8890013} are trying to detect the 21 cm brightness temperature fluctuation.
%\cite{Mellema:2006pd,PhysRevD.82.023006}.
The Experiment to Detect the Global EoR Signature (EDGES) could provide a lower limit 
on the width of reionization \cite{Monsalve:2016xbk}.

Another probe of reionization is the cosmic microwave background (CMB) 
\cite{Aghanim:1996ib,Gruzinov:1998un}. Reionization increases 
the number of free electrons, on which some of the CMB photons scatter.
This results in a suppression of the temperature and polarization angular power spectra at small scales, 
but reionization also serves as a source of polarization at the largest angular scales 
resulting in a very characteristic bump. 

Reionization is believed to be caused by the first appearance of luminous extreme UV ($\gtrsim 10$ eV) 
emitting objects such as heavy stars and quasars \cite{Couchman01071986,1990ApJ...350....1M}. 
The details of the astrophysical reionization at $z \sim 6$ are not yet well understood and are subject to 
analytical \cite{Madau:1998cd,Barkana:2000fd,Furlanetto:2004nh} and complicated numerical simulations \cite{Gnedin:1999fa,Ciardi:2003ia}.

Due to its dependence on the free electron fraction, the CMB is also sensitive to any other source of 
reionization, especially the decay or annihilation of dark matter (DM). This opens the very appealing 
possibility to obtain constraints on DM properties 
like its decay or annihilation rate. Various works have studied the impact of DM annihilation 
on the CMB, e.g.\  \cite{Natarajan:2008pk,Natarajan:2010dc,Chluba:2009uv,Giesen:2012rp,Lopez-Honorez:2013lcm,Poulin:2015pna,Ade:2015xua}, 
the list of literature on DM decay is in contrast surprisingly shorter, 
e.g.\ \cite{Kasuya:2003sm,Pierpaoli:2003rz,Kasuya:2006fq,Yeung:2012ya,Diamanti:2013bia}. 
Others also studied the astrophysical effects (not related to the CMB) of annihilating and 
decaying dark matter on the reionization history \cite{Iocco:2008xb,Freese:2008hb,Mapelli:2006ej,Liu:2016cnk}.

In this work, we study how model assumptions on astrophysical reionization
affect the inferred values of the amplitude of primordial scalar perturbations $A_{\text{s}}$, their spectral index 
$n_{\text{s}}$, and the DM decay rate $\Gamma$ from CMB temperature and polarization measurements. 

We derive constraints on the DM decay rate from the Planck 2015 data \cite{Ade:2015xua}. These constraints 
are in particular applicable for warm DM candidates such as keV sterile neutrinos \cite{Adhikari:2016bei}, which 
were recently reported to be observed with masses of $\sim 7$ keV \cite{0004-637X-789-1-13,Boyarsky:2014jta}. 
We thereby also include modifications to the CosmoRec code \cite{Chluba:2005uz,Chluba:2010ca} that 
are necessary to model DM decay realistically, but which we believe have been missed out in previous 
works \cite{Diamanti:2013bia,Lopez-Honorez:2013lcm}.  

Our work is organised as follows. In section \ref{Astrophysical reionization} we introduce two different 
parametrizations for the free electron fraction during reionization by means of astrophysical processes
-- the parametrization used in the CAMB code and a new parametrization proposed by \cite{Douspis:2015nca} 
based on observations of quasars and star forming galaxies. In section \ref{Dark matter decay and 
cosmic reionization} we discuss the decay of DM as an alternative source of reionization and its 
potential impact on the CMB. We present and discuss the results of our cosmological data analysis in 
section \ref{Comparison with CMB data} and conclude in section \ref{Conclusions}. Details on the 
numerical implementation can be found in the appendix \ref{appendixA} and  
an aspect of the prior distributions of the Bayesian inference problem are discussed in appendix \ref{appendixB}.

%%%%%%%%%%%%%%%%%%%%%%%%%%%%%%%%%%%%%%%%%%%%%%%%%%%%%%%%%%
\section{Astrophysical reionization}
\label{Astrophysical reionization}

\subsection{Evolution of the free electron fraction}
\label{Evolution of the free electron fraction}

\begin{figure}
\centering
\includegraphics[width=0.6\textwidth]{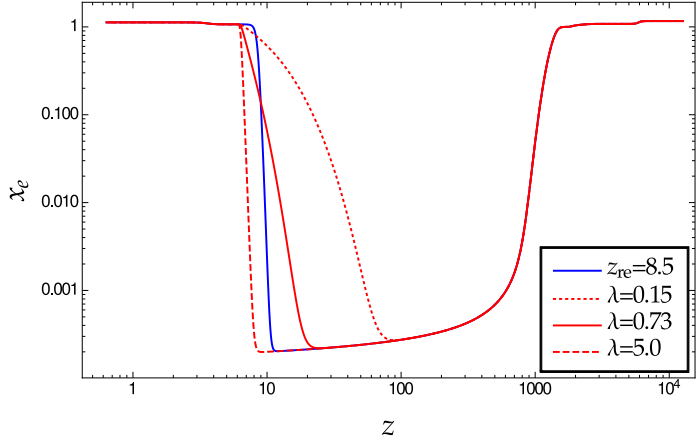}
\caption{Evolution of the free electron fraction $x_e$ for different models of 
astrophysical reionization. At high redshifts we show the fraction of free electrons as obtained by means of the 
CosmoRec code. For intermediate redshifts the blue line shows the parametrization \eqref{xe_standard} as 
used by the CAMB code and in the Planck 2015 data analysis (blue line). It is compared to the 
empirical parametrization \eqref{xe_alternative} (red lines) for three different values of the model 
parameter $\lambda$.}
\label{xe_high}
\end{figure}

The evolution of the free electron fraction $x_{\text{e}}=n_{\text{e}}/n_{\text{H}}$ in an isotropic and homogenous 
universe is calculated by recombination codes like CosmoRec \cite{Chluba:2005uz,Chluba:2010ca} that take 
into account multi-level excitations of hydrogen besides radiative transfer corrections of its recombination rate. 
Whereas helium recombines to HeII at $z \simeq 4000$ and to HeI at $z \simeq 1000$, hydrogen remains 
ionized until $z \simeq 900$. Due to the decrease of temperature its recombination 
rate finally exceeds the photoionization rate of excited hydrogen atoms and $x_{\text{e}}$ rapidly decreases 
to a residual value of the order $10^{-4}$, see figure~\ref{xe_high}. Afterwards, the free electron 
fraction is believed to smoothly continue to decrease until the onset of reionization by astrophysical sources 
at $z \sim 10$. 

For the purpose of CMB analysis the history of astrophysical reionization can be para\-metrized by a 
model for the evolution of the free electron fraction. Since the details of the astrophysical 
processes that cause cosmic reionization are still widely unknown, CAMB models $x_{\text{e}}(z)$ 
during astrophysical reionization as a smooth, step-like function, 
\begin{equation}
x_{\text{e}}(z) \big \vert_{\text{CAMB}}= 
\frac{1.08}{2} \left[1+ \tanh \left( \frac{(1+z_{\mathrm{re}})^{3/2}-(1+z)^{3/2}}{\Delta_z} \right) \right].
\label{xe_standard}
\end{equation}
Here, $z_{\mathrm{re}}$ describes the redshift of reionization, i.e. the redshift at which half of the electrons 
are free and $\Delta_z$ quantifies the width of the transition. Typically, cosmological data analysis uses 
$\Delta_z=0.5$ per default.

\begin{figure}
      \centering
		\includegraphics[width=0.6\textwidth]{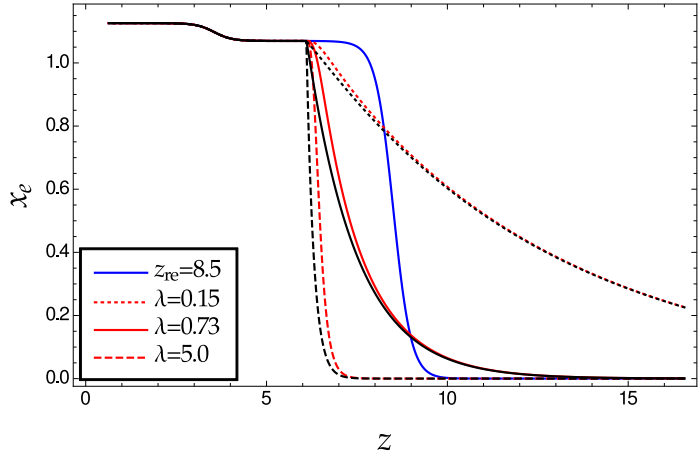}	
\caption{The free electron fraction during the epoch of reionization. 
The CAMB parametrization \eqref{xe_standard} (blue line) is compared to the empirical parametrization 
\eqref{xe_alternative} (red lines). The black lines describe the sharp edge introduced in 
\eqref{Q}, whereas the red curves show a smoothed version used in this work, see appendix \ref{appendixA} for details.}
\label{xe_comparison}
\end{figure}

By defining the free electron fraction as $x_{\text{e}}\equiv n_{\text{e}}/n_{\text{H}}$ the enumerator counts \textit{all} free electrons, whereas the denominator only counts hydrogen nuclei. Therefore, $x_{\text{e}}$ takes the asymptotic value of $\sim 1.08$ assuming 
full ionization of hydrogen plus single ionization of helium and $\sim 1.16$ assuming additionally 
double ionization of helium,
\begin{equation}
\begin{aligned}
x_{\text{e}} & \equiv \frac{n_{\text{e}}}{n_{\text{H}}} = \frac{n_{\text{e,HII}}}{n_{\text{H}}}+\frac{n_{\text{e,HeII}}}{n_{\text{H}}}+\frac{n_{\text{e,HeIII}}}{n_{\text{H}}} \\
& \xrightarrow{\text{compl. reion.}} \, 1.0 + 0.08 + 0.08 \,.
\label{xe_def}
\end{aligned}
\end{equation}
Since normal stellar populations produce only very few photons with energies above $54.4$ eV, the second ionization of helium is believed to be caused by the appearance of quasars at lower redshifts and is therefore included in CAMB by a second step at $z=3.5$, see figure~\ref{xe_comparison}. 

An alternative parametrization of reionization has recently been proposed in \cite{Douspis:2015nca}. Based on Lyman-$\alpha$ emission of star forming galaxies and Gunn-Peterson optical depths of quasars \cite{Bouwens:2015vha}, the authors propose the following asymptotic behaviour of the ionised hydrogen fraction $Q_{\text{HII}}=n_{\text{HII}}/n_{\text{H}}$
\begin{equation}
\begin{aligned}
1-Q_{\text{HII}} & \propto (1+z)^3 \, && \text{for } z < z_{\text{p}} \\
Q_{\text{HII}} & \propto \exp (-\lambda(1+z)) \, && \text{for } z \geq z_{\text{p}}.
\end{aligned}
\label{QHII_asymptotic}
\end{equation}
Such an empirical parametrization is also corroborated by numerical simulations that show extended 
scenarios of reionization \cite{2041-8205-756-1-L16,Park:2013mv}.
As proposed in \cite{Douspis:2015nca}, this parametrization can be characterized by three free parameters: the 
pivot redshift $z_{\text{p}}$, the ionised hydrogen fraction at the pivot scale $Q_{\text{p}}=Q_{\text{HII}}(z_{\text
{p}})$ and the evolution parameter in the exponential $\lambda$. 
This translates into the following parametrization of $Q_{\text{HII}}$
\begin{equation}
Q_{\text{HII}} (z) = 
\begin{cases} \frac{1-Q_{\text{p}}}{(1+z_{\text{p}})^3-1} \left( (1+z_{\text{p}})^3-(1+z)^3 \right) +
Q_{\text{p}} \hspace{1cm} & \text{for } z < z_{\text{p}}\\
Q_{\text{p}}  e^{-\lambda(z-z_{\text{p}})} & \text{for } z \geq z_{\text{p}}. \end{cases} 
\label{Q}
\end{equation}

We furthermore assume that $Q_{\text{HII}}(z=0) =1$, in order to fix the last degree of freedom 
that is allowed by \eqref{QHII_asymptotic}. According to \cite{Douspis:2015nca} the best-fit parameters are: 
$z_{\text{p}}=6.1$, $Q_{\text{p}}=0.99986$ and $\lambda=0.73$. Note that these fits are based on 
\textit{direct observations} of $Q_{\text{HII}}$.

In order to compare this parametrization to the one used by CAMB,  \eqref{xe_standard}, we first have to find 
an expression for the free electron fraction $x_{\text{e}}$, as defined in \eqref{xe_def}. Since $Q_{\text{HII}}$ only 
accounts for the first term in eq. \eqref{xe_def}, we have to make an additional assumption about the ionization 
of helium. Considering the relatively similar ionization energies of 13.6 eV for HI and 24.6 eV for HeI, it seems 
reasonable to assume that the 
first ionization of helium happens in the same manner as the ionization of hydrogen. In that case, we can 
simply write
\begin{equation}
x_{\text{e}}(z)\big \vert_{\text{emp.}} = 1.08 \times Q_{\text{HII}} (z)
\label{xe_alternative}
\end{equation}
and the second ionization is described by a second step at $z=3.5$, analogously as in the CAMB 
parametrization \eqref{xe_standard}, see figure~\ref{xe_comparison}.

For $z < z_{\text{p}}$, $Q_{\text{HII}}$ is tightly constrained by astrophysical observations \cite{Bouwens:2015vha}. Therefore, in the following
we fix $Q_{\text{p}}$ and $z_{\text{p}}$ to the best-fit values quoted above, and keep $\lambda$ as the single 
free parameter. We refer to the parametrization \eqref{xe_alternative} as \textit{empirical parametrization} and 
to \eqref{xe_standard} as \textit{CAMB parametrization} in the rest of this work.

In figures \ref{xe_high} and \ref{xe_comparison} we show the free electron fraction as a function of redshift for 
the empirical parametrization \eqref{xe_alternative} for three different values of $\lambda$ compared to the 
CAMB parametrization \eqref{xe_standard} with $z_{\text{reio}}=8.5$. The empirical parametrization with 
small values of $\lambda$ allows for more extended reionization histories than the CAMB parametrization 
with $\Delta_z=0.5$. 

Equation \eqref{Q} exhibits a sharp edge at $z_{\text{p}}=6.1$ and therefore leads to a discontinuity in 
the derivatives of the visibility function. We therefore have to smooth $x_{\text{e}}$ \eqref{xe_alternative} 
at $z_{\text{p}}$ in order to remove unphysical bias. Details can be found in the appendix \ref{appendixA}. We 
include a plot of the smoothed version of eq.~\eqref{Q} in figure~\ref{xe_comparison}.

\subsection{Impact of astrophysical reionization on CMB angular power spectra}
\label{Impact of reionization on CMB angular power spectra}

Reionization affects the high-$\ell$ and low-$\ell$ CMB spectra in two ways:
\begin{itemize}
\item[i)] The CMB temperature fluctuations experience a suppression of $e^{-\tau}$, where $\tau$ denotes the optical depth defined as
\begin{equation}
\tau(z) \equiv c \int_z^0 n_e(z') \sigma_{\text{T}} \frac{\mathrm{d}t}{\mathrm{d}z'} \mathrm{d}z'.
\label{optical_depth}
\end{equation}
Above $t$ denotes cosmic time and $\sigma_{\text{T}}$ is the Thomson scattering cross section. 
The temperature angular power spectrum (TT) is therefore suppressed by a factor of 
$e^{-2\tau}$. Thus the optical depth $\tau$ is degenerate with the amplitude of 
primordial curvature fluctuations, described by $A_{\text{s}}$ (we do not consider primordial 
tensorial fluctuations in this work). The suppression is present at all scales that are subhorizon before 
recombination, i.e.~$\ell \gtrsim 200$. For lower $\ell$ the suppression of the spectrum is
less pronounced, as these modes enter the horizon after recombination or even after reionization 
($\ell \lesssim 20$).
\item[ii)] The polarization angular power spectra (parity-even E-modes and parity odd B-modes, the 
latter are not further studied in this work) reflect the size of the temperature 
quadrupole at the time of last scattering at a given scale. The polarization and cross 
angular power spectra (TE, TB, EE, EB, BB) are also suppressed by a factor $e^{-2 \tau}$. But more 
importantly, reionization also causes polarization at large scales, where without reionization there 
would be no polarization source due to the lack of electrons that could scatter the intensity quadrupole.
This effect shows up as a bump in the polarization angular power spectra at low $\ell$ and is a unique 
feature of reionization as it can only be caused by a late time effect.
\end{itemize} 
 
\begin{figure}
\centering
	\begin{subfigure}[b]{0.6\textwidth}
		\includegraphics[width=\textwidth]{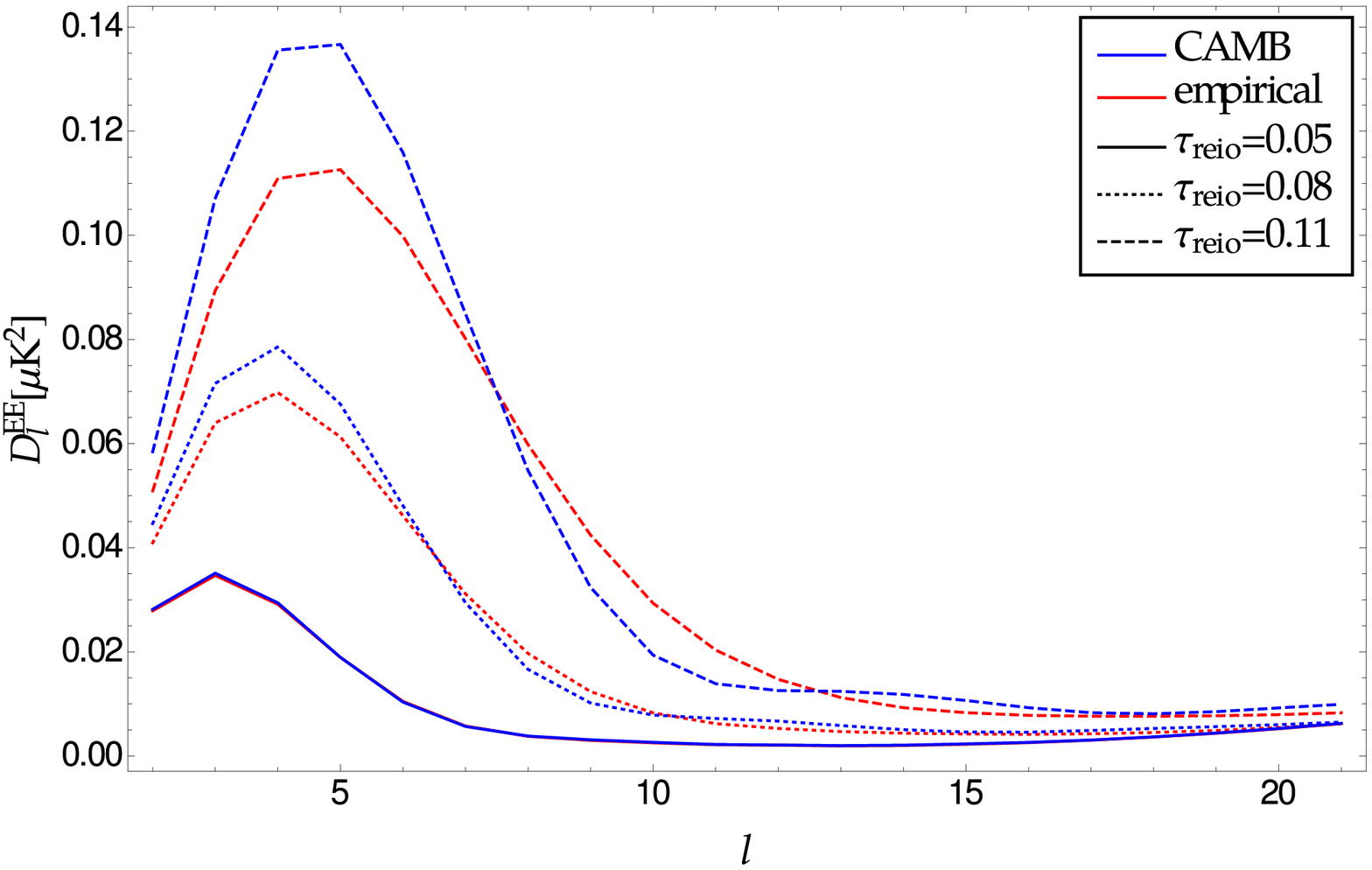}
	\end{subfigure}
	\begin{subfigure}[b]{0.6\textwidth}
		\includegraphics[width=\textwidth]{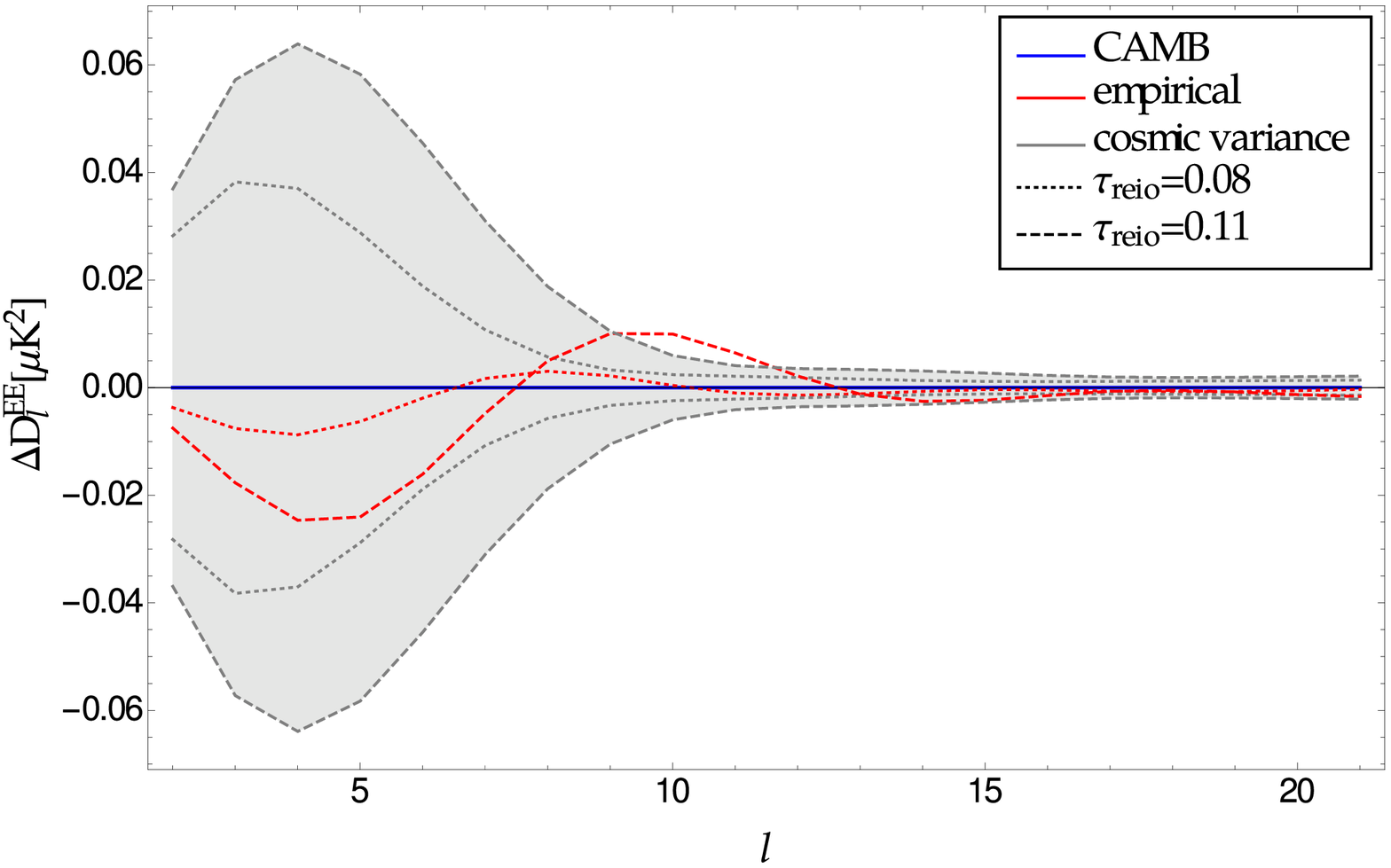}
	\end{subfigure}		
\caption{Top panel: Comparison of the CMB EE angular band power spectra 
($D^{EE}_{\ell}=\ell(2\ell +1) C^{EE}_{\ell} /2 \pi$) for the CAMB \eqref{xe_standard} (blue lines) and 
empirical parametrizations \eqref{xe_alternative} (red lines) for three different values of the optical depth. 
Bottom panel: Difference 
$\Delta D^{EE}_{\ell} = D^{EE}_{\ell,\text{emp}} - D^{EE}_{\ell,\text{CAMB}}$ in the EE spectrum 
(red lines) at equal optical depth. The shaded regions represent the cosmic variance for the respective 
values of the optical depth.}
\label{Cl_tau}	
\end{figure}

Therefore, the high-$\ell$ TT and EE spectra only depend on the amount of reionization, but not on the details 
of the reionization history itself. There are however two important caveats to this simplistic picture: Firstly and 
most importantly for the scenarios considered in this work, the low-$\ell$ data have in principle the potential 
to distinguish between different reionization histories. This can be seen in figure \ref{Cl_tau}, where we show the 
low-$\ell$ EE spectrum for the CAMB reionization \eqref{xe_standard} and the empirical reionization 
\eqref{xe_alternative} for the same values of the optical depth. The CAMB parametrization gives rise to
more polarization power at low $\ell$ than the empirical parametrization. This difference is more pronounced 
for high values of $\tau$ and vanishes for small $\tau$. 
It is also important to ask how the difference between reionization histories compares to cosmic 
variance of the polarization power spectra. This is shown in the lower figure \ref{Cl_tau}. We notice that the 
differences are less significant at very small $\ell$, but can be larger than the cosmic variance for individual 
multipole moments at $\ell \sim 10$ and above.  
See also \cite{Kaplinghat:2002vt}, 
\cite{Natarajan:2010dc} or \cite{Poulin:2015pna} for other examples of different reionization histories and 
their impact in the low-$\ell$ polarization spectrum.

Secondly, as shown in \cite{Poulin:2015pna,Giesen:2012rp}, for significant changes in the reionization history at 
$\textit{early}$ times, the suppression of the high-$\ell$ CMB spectra becomes oscillatory and the simple picture 
described above breaks down. This effect stems from substantial changes in the visibility function at early times, 
but turns out to be irrelevant for all scenarios considered in this work.

\begin{figure}
\centering
        \begin{subfigure}[b]{0.49\textwidth}
                \includegraphics[width=\textwidth]{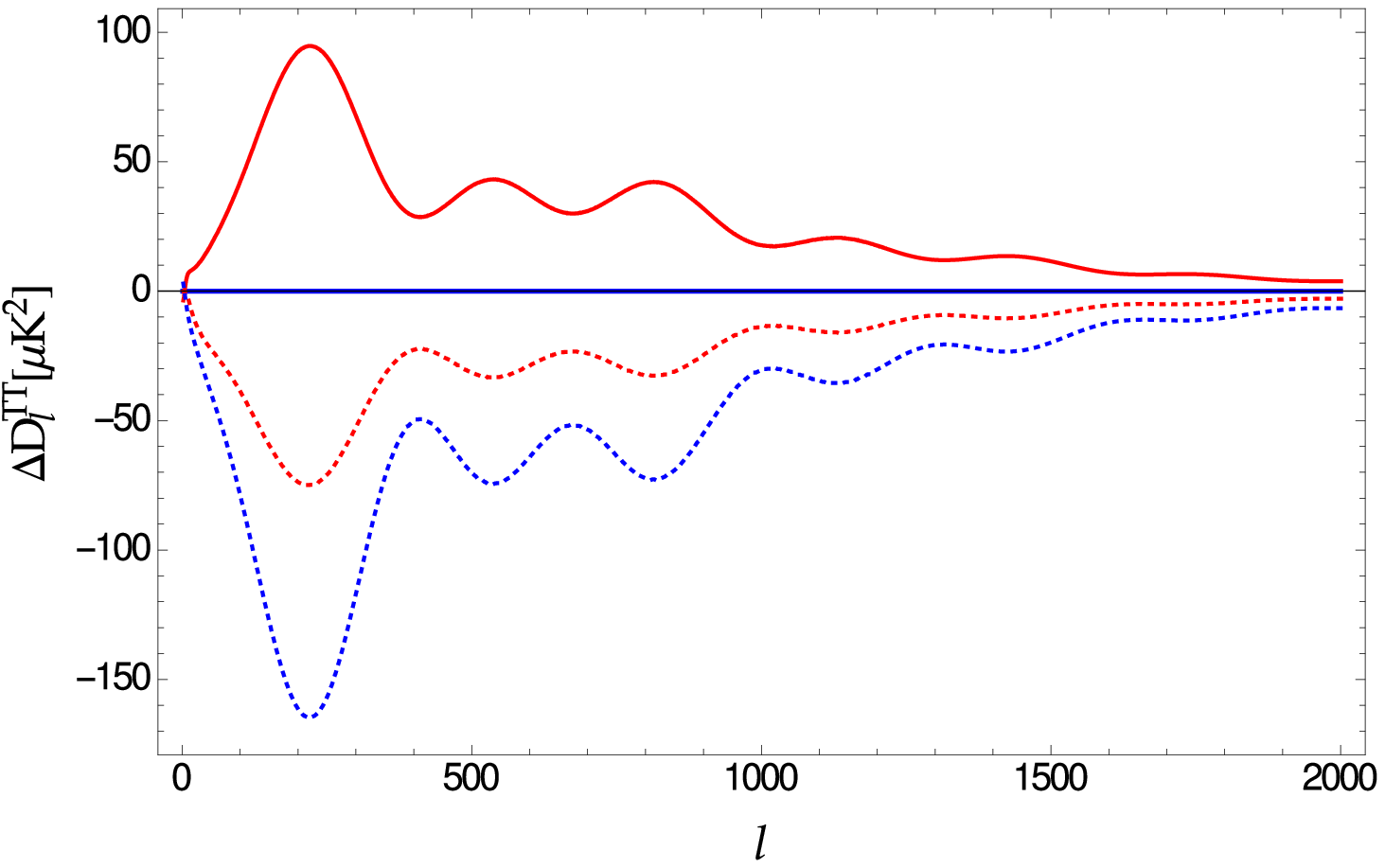}
        \end{subfigure}
        \begin{subfigure}[b]{0.49\textwidth}
                \includegraphics[width=\textwidth]{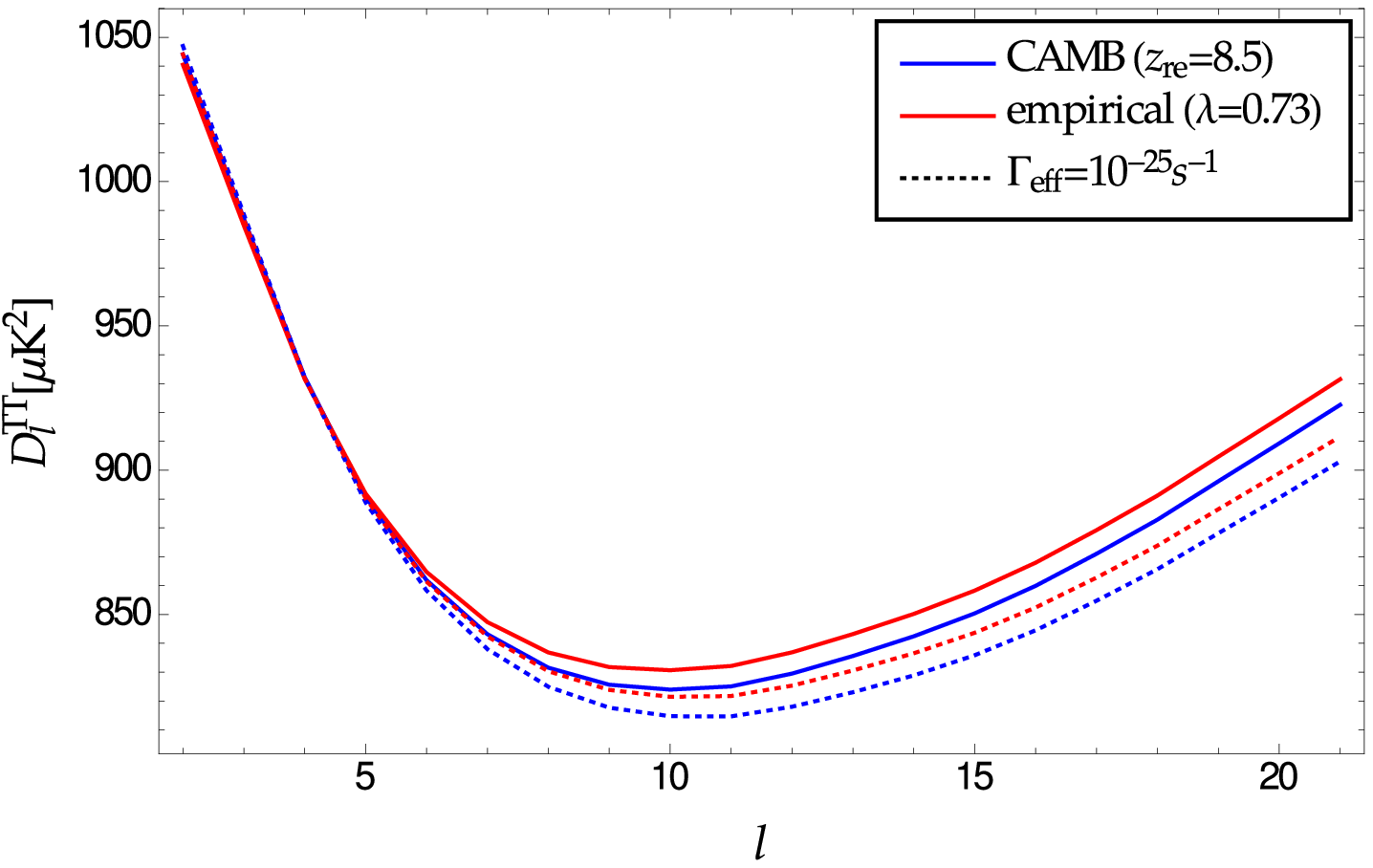}
        \end{subfigure}
        \begin{subfigure}[b]{0.49\textwidth}
                \includegraphics[width=\textwidth]{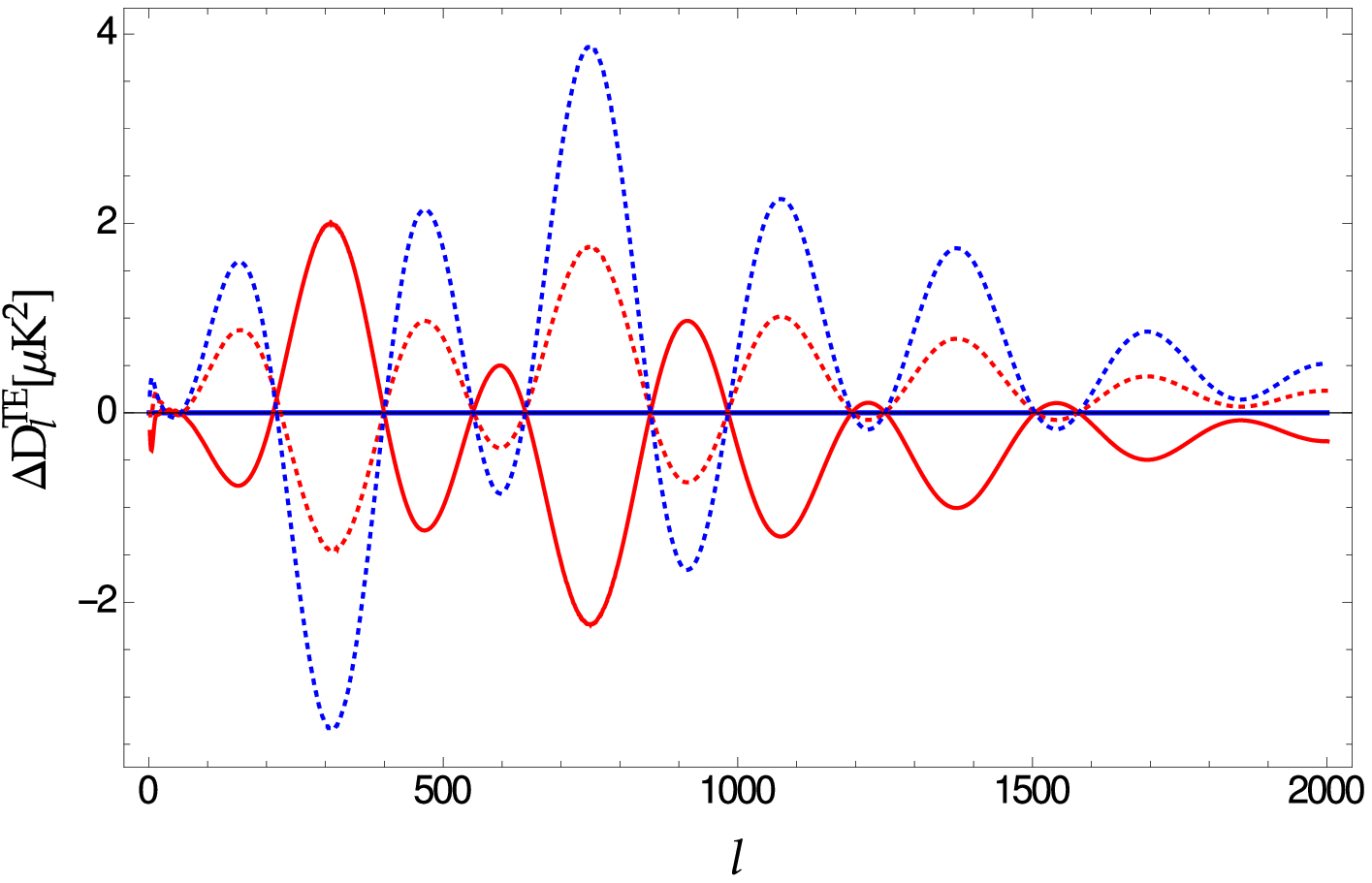}
        \end{subfigure}
        \begin{subfigure}[b]{0.49\textwidth}
                \includegraphics[width=\textwidth]{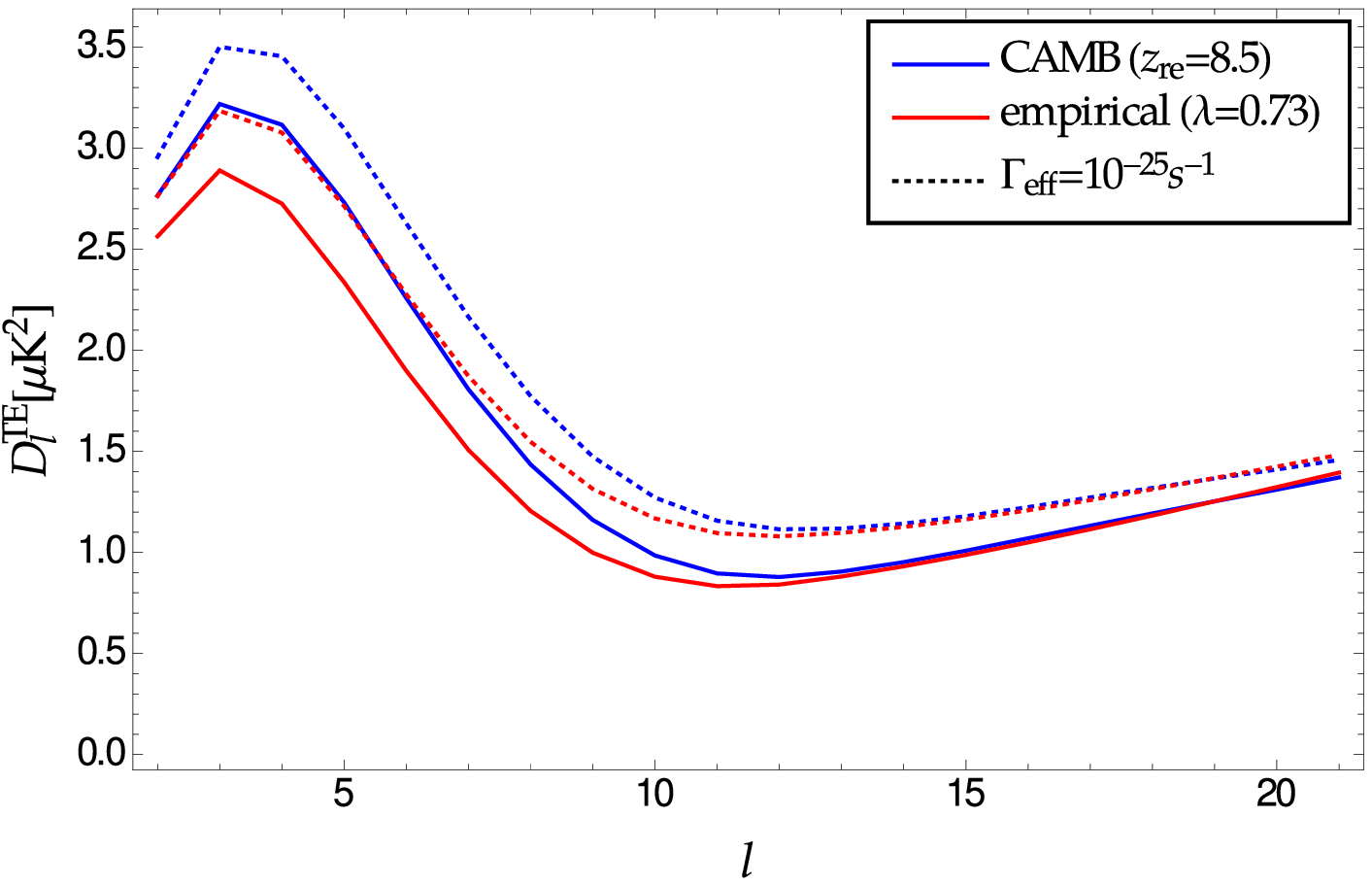}
        \end{subfigure}
        \begin{subfigure}[b]{0.49\textwidth}
                \includegraphics[width=\textwidth]{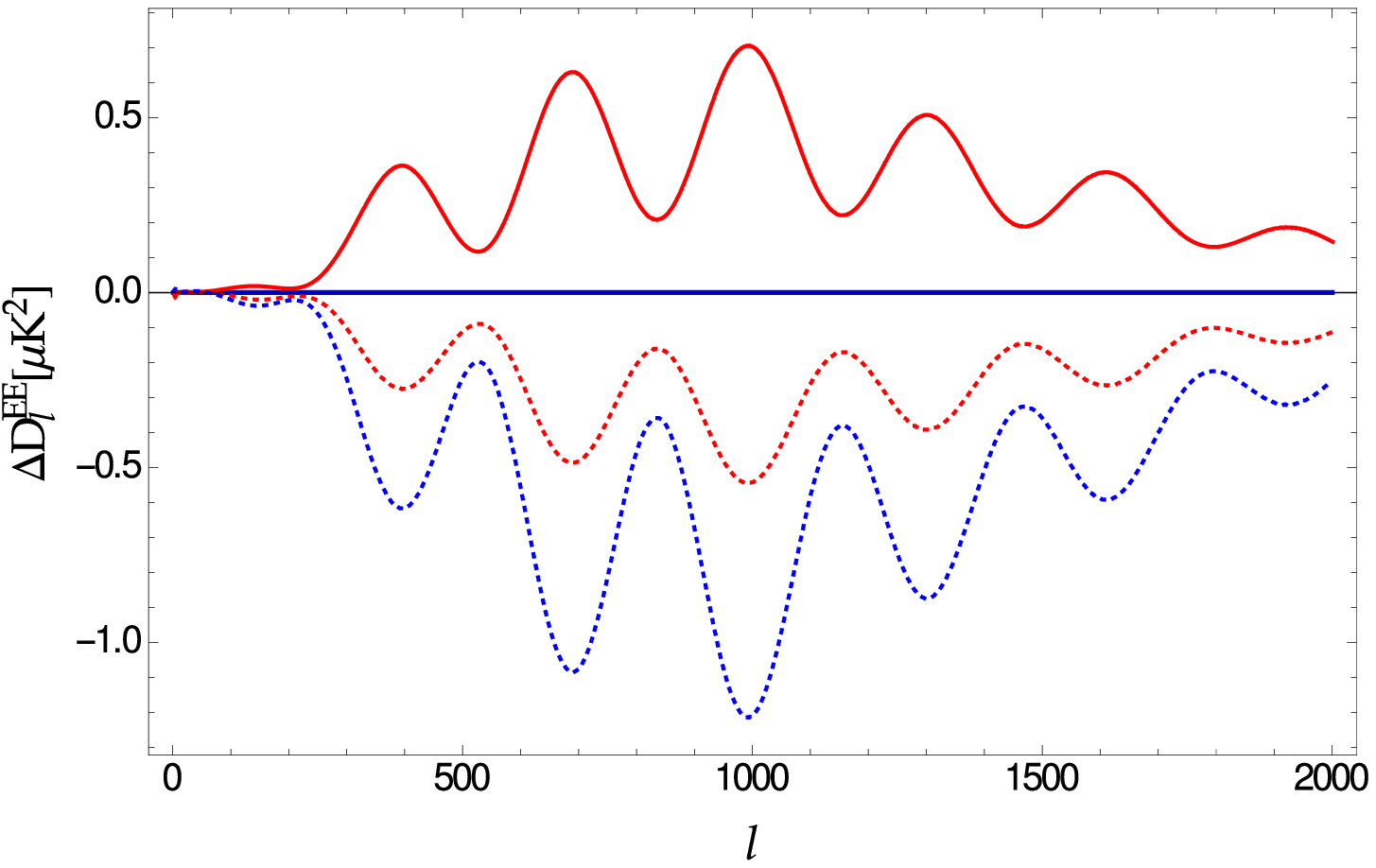}
        \end{subfigure}
        \begin{subfigure}[b]{0.49\textwidth}
                \includegraphics[width=\textwidth]{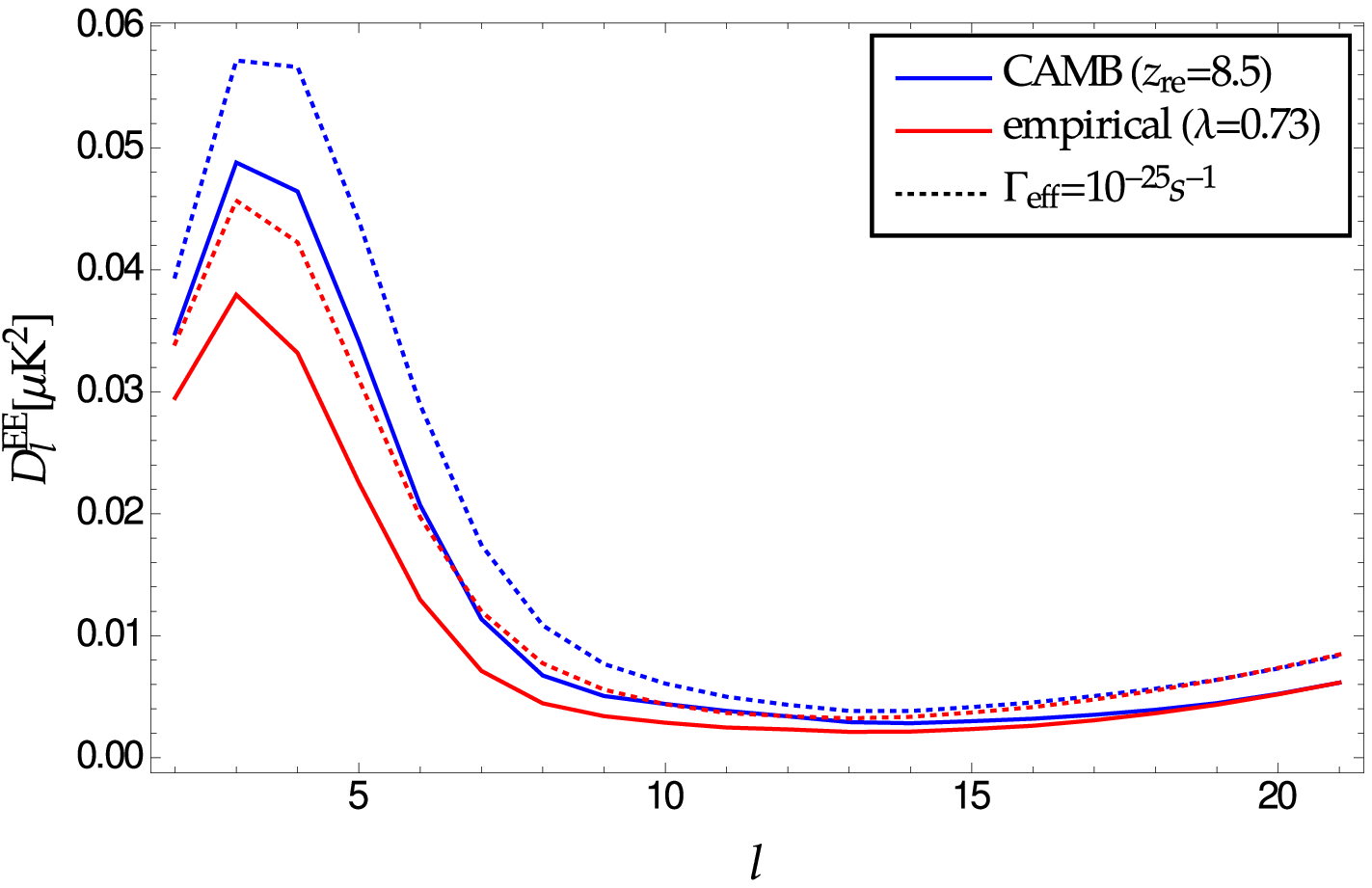}
        \end{subfigure}
\caption{CMB angular band power spectra (top row:\ TT, middle row:\ TE, bottom row:\ EE) for the 
CAMB parametrization \eqref{xe_standard} with $z_{\text{re}}=8.5$ (blue lines), the empirical parametrization 
\eqref{xe_alternative} with $\lambda=0.73$ (red lines), both with (dashed lines) and without (solid lines) 
dark matter decay for $\Gamma_{\text{eff}}=10^{-25}s^{-1}$. The left column shows the angular band 
power difference to the CAMB parametrization ($z_{\text{re}}=8.5$) without dark matter decay, 
$\Delta D_{\ell}=D_{\ell,i}-D_{\ell,\text{CAMB}}$. The right column shows the band spectra in 
the so-called reionization bump region at low multipole moments $\ell < 20$.}
\label{CMB_spectrum}	
\end{figure}

In figure~\ref{CMB_spectrum}, we show the CMB TT, EE and TE angular power spectra for the 
CAMB parametrization \eqref{xe_standard} with the Planck 2015 best-fit value $z_{\text{re}}=8.5$ ($\tau_{\text{reio}}$=0.061)
\cite{Ade:2015xua} and the empirical parametrization \eqref{xe_alternative} for the best-fit 
value $\lambda=0.73$ ($\tau_{\text{reio}}$=0.053) \cite{Douspis:2015nca}. The figure also includes the effect of DM decay which we discuss later in section \ref{Impact of dark matter decay on the CMB angular power spectra}.  

%%%%%%%%%%%%%%%%%%%%%%%%%%%%%%%%%%%%%%%%%%%%%%%%%%%%%%%%%%
\section{Dark matter decay and cosmic reionization}
\label{Dark matter decay and cosmic reionization}

\subsection{Dark matter decay}
\label{Dark matter decay}

Let us turn to study the impact of an additional source of reionization, namely the decay of dark matter. 
To avoid any source of confusion, we will refer to the late time reionization by astrophysical objects, 
i.e.~\eqref{xe_standard} or \eqref{xe_alternative}, by \textit{astrophysical reionization} and to 
reionization due to DM decay or annihilation by \textit{DM reionization}. 

Various works investigated such scenarios of DM reionization, e.g.~\cite{Natarajan:2008pk,Natarajan:2010dc,Chluba:2009uv,Giesen:2012rp,Lopez-Honorez:2013lcm,Diamanti:2013bia,Kasuya:2003sm,Pierpaoli:2003rz,Kasuya:2006fq,Yeung:2012ya}, 
most of them focussing on DM annihilations. But all of these studies are based on the CAMB parametrization 
of astrophysical reionization \eqref{xe_standard}. Only recently, the authors of \cite{Poulin:2015pna} have
investigated the impact of DM halo formation on the the reionization history, considering thereby for the first 
time \textit{two different astrophysical reionization histories}. They came to the conclusion that in the case of 
the CAMB parametrization \eqref{xe_standard}, the influence of halo formation is negligible, whereas for 
an alternative parametrization based on the star formation rate, the impact of halo formation can 
become substantial. This immediately calls for a closer investigation of the impact of our assumptions 
about astrophysical reionization on constraints of DM properties. In the following, we start with a description 
of the impact of DM decay to the reionization history.

The evolution of the number density $n_d$ of the decaying DM particles is described by the equation 
of radioactive decay in an expanding universe,
\begin{equation}
\dot{n}_{\text{d}} (t) +3 H n_{\text{d}}(t)=-\Gamma_{\text{tot}} n_{\text{d}} (t).
\label{Boltzmann}
\end{equation}
Here $\Gamma_{\text{tot}}= \Gamma_{\text{em,s}}+ \Gamma_{\text{w}}$ is the sum of the decay rate into 
particles only interacting via weak force (neutrinos), $\Gamma_{\text{w}}$, and the decay rate into 
particles interacting via electromagnetic and/or strong force, $\Gamma_{\text{em,s}}$. 
We refer to this decaying particle as DM, even though we do not restrict to the case where it constitutes all 
DM in the universe. The solution of equation \eqref{Boltzmann} is given by
\begin{equation}
n_{\text{d}}=n_{\text{d,i}} \left( \frac{a_i}{a} \right)^3 e^{-\Gamma_{\text{tot}} (t-t_i)} = 
n_{\text{d,i}} \left( \frac{1+z}{1+z_i} \right)^3 \exp\left(- \Gamma_{\text{tot}} 
\int_z ^{z_{\text{i}}} \frac{\text{d} z}{(1+z)H(z)} \right), 
\end{equation}
where $n_{\text{d,i}}$ is the initial number density. For low decay rates $\Gamma_{\text{tot}}$ or at 
sufficiently early times $t$, we find $n_{\text{d}} \propto a^{-3}$, i.e.~the dominant dilution effect is 
due to the expansion of the universe.

\subsection{Energy deposition from dark matter decay}
\label{Energy deposition from dark matter decay}

The decaying DM injects energy into the cosmic hydrogen-helium gas. 
The energy that is injected per time and volume via electromagnetically or strongly interacting particles 
into the medium by the decay is given by
\begin{equation}
\left( \frac{dE}{dt \, dV} \right)_{\text{inj}} = \Delta E_{\text{d}} \, \Gamma_{\text{em,s}} n_{\text{d}}(t).
\label{E_inj}
\end{equation}
To indicate that we also include the scenario for a DM particle going from an excited to some lower state, 
we have written $\Delta E_{\text{d}}$ in \eqref{E_inj}. In such a scenario, $n_{\text{d}}$ also has to be 
interpreted as the number density of the excited state. For the case of DM decay into standard model 
particles, we have $\Delta E_{\text{d}}=m_{\text{d}}$, where $m_{\text{d}}$ is the DM mass.

In general, not all of the injected energy is immediately deposited into the surrounding medium. If the 
density of the neighbouring gas is low, the emitted particles experience redshift before they are absorbed. 
This effect depends on the redshift of consideration, on the injected energy $\Delta E_{\text{d}}$ 
as well as the nature of the emitted particles. The total effect is usually very hard to compute, because it 
includes the formation of cascades and their efficiency to ionize the gas. Moreover, decay products in the 
form of neutrinos lead to further energy loss, because neutrinos simply free-stream and do not interact 
with the medium. In general, the relation between injected and emitted energy is given by
\begin{equation}
\left( \frac{dE}{dt \, dV} \right)_{\text{dep}} \!\!\!\!\! (z)= \int_{z}^\infty F_{\text{m}}  (z,z_{\text{inj}},\Delta E_{\text{d}}) \left( \frac{dE}{dt \, dV} \right)_{\text{inj}} \!\!\!\!\! (z_{\text{inj}}) \, \, \, \mathrm{d}z_{\text{inj}} ,
\label{E_dep1}
\end{equation}
where the function $F_{\text{m}}(z,z_{\text{inj}}, \Delta E_{\text{d}})$ describes the fraction of the energy injected at redshift $z_{\text{inj}}$ which is deposited at redshift $z$. The label m indicates the dependence on the specific model of DM decay. However, it is a common practice to include this effect by a function $f_{\text{m}}(z,\Delta E_{\text{d}})$ that encapsules the above integral, which in combination with eq.~\eqref{E_inj} gives
\begin{eqnarray}
\left[ \frac{dE}{dt \, dV} \right]_{\text{dep}} \!\!\!\!\!\!\! (z) &= 
& \left[\frac{1}{\left[ \frac{dE}{dt \, dV} \right]_{\text{inj}}\! (z)}\int_{z}^\infty 
F_{\text{m}}  (z,z_{\text{inj}},\Delta E_{\text{d}}) \left[ \frac{dE}{dt \, dV} \right]_{\text{inj}} \!\!\!\!\! (z_{\text{inj}}) \, \, \, \mathrm{d}z_{\text{inj}}\right] \left[ \frac{dE}{dt \, dV} \right]_{\text{inj}}\!\!\!\!\!\!(z) \nonumber \\ 
& \equiv & f_{\text{m}}(z,\Delta E_{\text{d}})\left( \frac{dE}{dt \, dV} \right)_{\text{inj}}\!\!\!\!\!(z) \nonumber \\ 
& = & \frac{m_{\text{p}}}{0.76} \,n_{\text{H}}(z) f_{\text{m}}(z,\Delta E_{\text{d}}) \left( \frac{\Delta E_{\text{d}}}{m_{\text{d}}} \right) \left( \frac{\rho_{\text{d}}}{\rho_{\text{b}}} \right) \Gamma_{\text{em,s}}  \nonumber \\
&  = & 1.23 \times 10^9 \, n_{\text{H}}(z)  \, f_{\text{m}}(z,\Delta E_{\text{d}}) \left( \frac{\Delta E_{\text{d}}}{m_{\text{d}}} \right) \left( \frac{\rho_{\text{d}}}{\rho_{\text{b}}} \right) \Gamma_{\text{em,s}} \, \, \, \text{eV} \,,
\label{E_dep}
\end{eqnarray}
where $n_{\text{H}}$ is the number density of hydrogen and we use $n_{\text{H}}=0.76 n_{\text{b}}$ in the second line. 

From now on we focus on cases where $(\rho_{\text{d}}/ \rho_{\text{b}}) \simeq$ const, i.e.~lifetimes much 
larger than the age of the universe or $\Gamma_{\text{tot}} \lesssim 10^{-17}s^{-1}$. Inclusion of higher 
decay rates would furthermore demand the modification of the Friedmann equations to take into 
account the decay of non-relativistic matter into radiation \cite{Poulin:2016nat,Audren:2014bca,Lattanzi:2013uza}. If the decaying particle makes up all dark matter 
of the universe we have \mbox{$(\rho_{\text{d}}/ \rho_{\text{b}}) \approx 5.5$}, but we keep the 
possibility of $0<(\rho_{\text{d}}/ \rho_{\text{b}}) \lesssim 5.5$. Since eq. \eqref{E_dep} depends in the 
same way on $(\rho_{\text{d}}/ \rho_{\text{b}})$, $(\Delta E_{\text{d}}/ m_{\text{d}})$ and 
$\Gamma_{\text{em,s}}$, it is convenient to summarize these three parameters as
\begin{equation}
\Gamma_{\text{eff}}= \left( \frac{\Delta E_{\text{d}}}{m_{\text{d}}} \right) \, \left( \frac{\rho_{\text{d}}}{\rho_{\text{b}}} \right) \Gamma_{\text{em,s}} 
\label{Gamma'}
\end{equation}
and therefore
\begin{equation}
\left( \frac{dE}{dt \, dV} \right)_{dep}= 1.23 \times 10^9 \, n_{\text{H}}  \, f_{\text{m}}(z,\Delta E_{\text{d}}) \, \Gamma_{\text{eff}} \, \, \, \text{eV}.
\label{E_dep2}
\end{equation}

\subsubsection{On-the-spot approximation}
\label{On-the-spot approximation}

The function $f_{\text{m}}(z,\Delta E_{\text{d}})$ can for example be computed for different $\Delta E_{\text{d}}$ with the publicly available code described in \cite{Slatyer:2015kla}. Taking into account the full possible energy range of $\Delta E_{\text{d}}$ into a cosmological data analysis is in general computationally  expensive. In section \ref{Comparison of the two parametrizations of astrophysical reionization} our aim is in the first place to investigate how much these constraints depend on our assumptions about astrophysical reionization. Therefore we restrict our analysis here to an energy range that allows us to use the so called \textit{on-the-spot} approximation in which $f_{\text{m}}(z,\Delta E_{\text{d}})=1$. This approximation restricts the validity of our analysis to $\Delta E_{\text{d}} \lesssim$ 1 keV \cite{Slatyer:2015kla,Slatyer:2009yq}. 

In section \ref{sec:wdm}, we apply our work to warm DM candidates such as keV-mass sterile neutrinos \cite{Adhikari:2016bei}. The specific parameters for this DM candidate can be constrained via a reinterpretation of the constraints on $\Gamma_{\text{eff}}$, as is shown in section \ref{sec:wdm}. Additionally to the on-the-spot approximation, we also consider the full redshift range of $f_{\text{m}}(z,\Delta E_{\text{d}})$ by using a fitting formula of the deposition efficiency obtained by \cite{Ripamonti:2006gq}. 

\subsection{Reionization from dark matter decay}
\label{Reionization from dark matter decay}

To find the evolution of the free electron function $x_{\text{e}}$ we have modified the recombination code
CosmoRec and included the new energy source term described in eq.~\eqref{E_dep}. As a further input we 
need to include how the energy \eqref{E_dep} is distributed, i.e.\ how much energy goes into ionization and 
excitation of atoms and how much into heating of the medium. The CosmoRec code already contains a 
module for DM annihilation and we assume that the DM decay has the same energy distribution, 
as described in \cite{Chluba:2009uv,Chen:2003gz}. As usual, the output of CosmoRec is included 
in the Boltzmann solver CAMB and then at some lower redshift the CAMB reionization module 
initiates the astrophysical reionization, i.e.\ \eqref{xe_standard} or \eqref{xe_alternative}.

We use the Recfast++ module of CosmoRec to model the effect of DM decay on the ionization history. This allows us to captures the main recombination physics corrections around $z \sim 1000$ (e.g., \cite{Shaw:2011ez}), while providing sufficient flexibility to account for large energy injection.

There are two further effects that need to be taken into account, as described in detail in 
\cite{Chluba:2015lpa} for the case of heating by primordial magnetic fields: 
\begin{itemize}
\item[i)] The photon ionization coefficient has to be evaluated as a function of photon temperature 
$T_{\gamma}$, not electron temperature $T_{\text{e}}$. Negligence of this can lead to a strong 
overestimation of the photon ionization rate which results in an extremely sudden increase of the 
free electron fraction.
\item[ii)] Collisional ionization effects become efficient for $T_{\text{e}} \gtrsim 10^4$K and therefore 
should be included in Recfast++. Large decay rates $\Gamma_{\text{eff}}$ lead to an enhancement 
of the ionization rate and hence to an increase of the free electron fraction.
\end{itemize}

\begin{figure}
\centering
	\begin{subfigure}[b]{0.6\textwidth}
		\includegraphics[width=\textwidth]{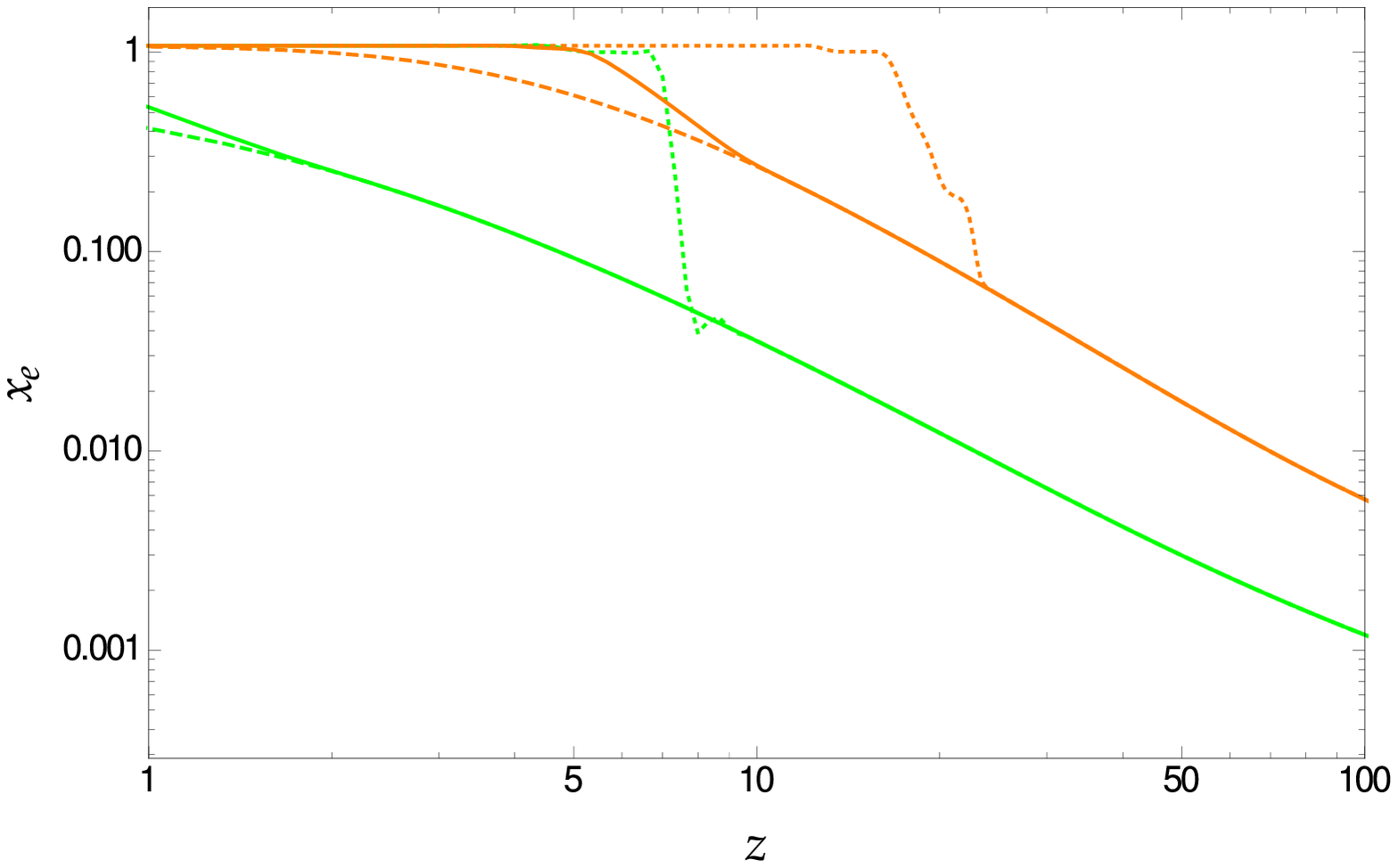}
	\end{subfigure}
	\begin{subfigure}[b]{0.6\textwidth}
		\includegraphics[width=\textwidth]{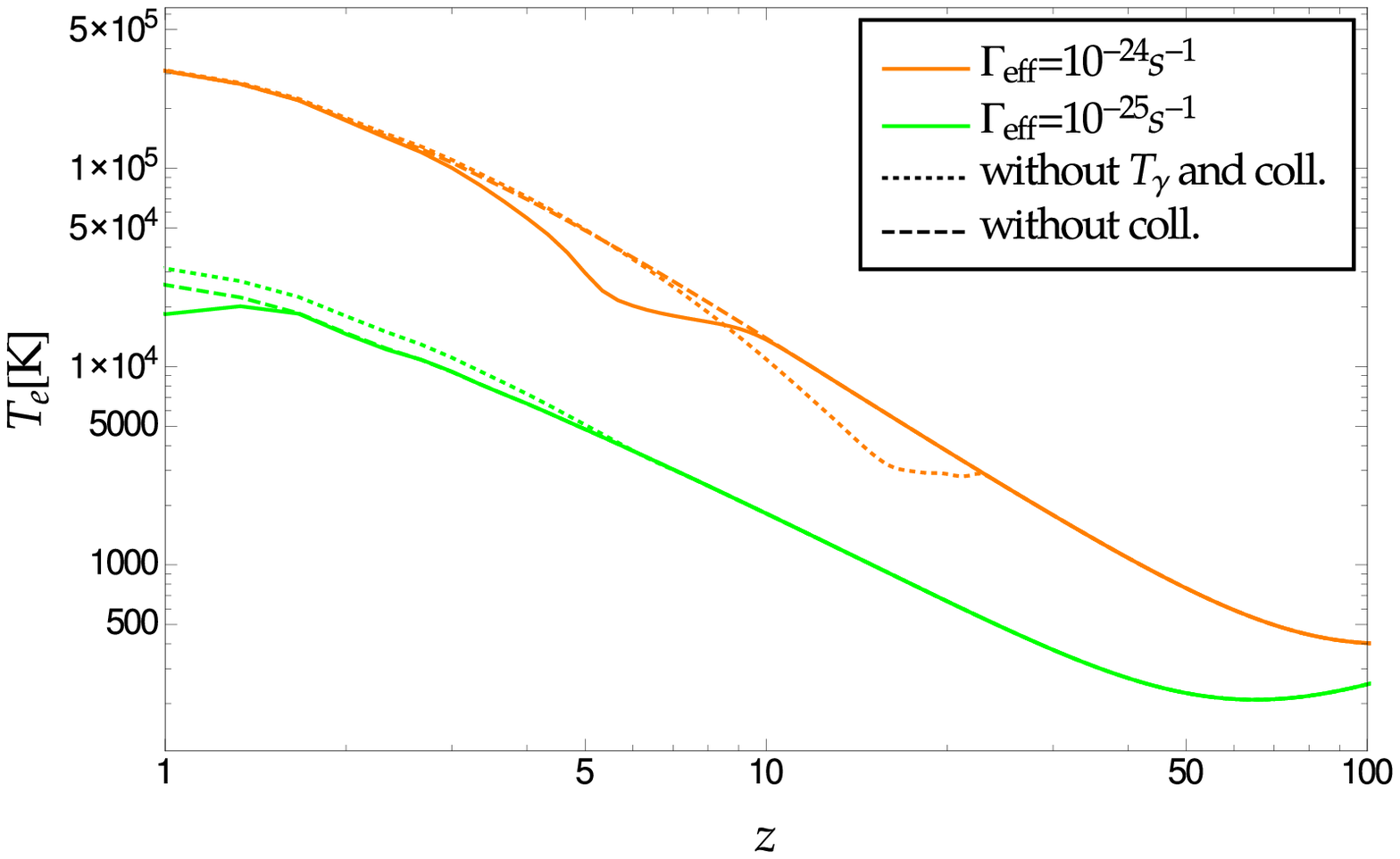}
	\end{subfigure}		
\caption{Cosmic reionization from dark matter decay. We plot the evolution of the free electron fraction 
(top panel) and the electron temperature (bottom panel) for $\Gamma_{\text{eff}}=10^{-25}$ s$^{-1}$ 
(green lines) and $\Gamma_{\text{eff}}=10^{-24}$ s$^{-1}$ (orange lines), ignoring contributions from 
astrophysical reionization. Dotted lines:\ The output produced by the public Recfast++ run mode.
Dashed lines:\ First code modification to evaluating the photon ionization coefficient at the photon 
temperature $T_{\gamma}$. Solid lines:\ Second code modification to include collisional ionizations.}
\label{CosmoRec}
\end{figure}

Both corrections are negligible for astrophysical recombination histories, but are very important for 
DM decay. In figure \ref{CosmoRec}, we show the impact of these corrections on the free electron 
fraction $x_{\text{e}}$ and the electron temperature $T_{\text{e}}$ for 
$\Gamma_{\text{eff}}=10^{-25}$ s$^{-1}$ and $\Gamma_{\text{eff}}=10^{-24}$ s$^{-1}$. 
Note that both plots only show the output of CosmoRec (in Recfast++ mode). The astrophysical 
reionization by CAMB, eq. \eqref{xe_standard} or \eqref{xe_alternative}, is \textit{not} yet added. Without both 
corrections (dotted line), the free electron fraction shows a very abrupt and implausible increase 
at $x_{\text{e}} \sim 0.05$. For both displayed $\Gamma_{\text{eff}}$ this unphysical transition 
happens at $z\geq 6$, i.e.\ possibly before the onset of 
astrophysical reionization, and leads to a serious overestimation of the impact of DM decay 
on the free electron fraction. We believe that earlier investigations, 
e.g.~\cite{Diamanti:2013bia,Lopez-Honorez:2013lcm}, have overlooked this effect. When correction 
i) is taken into account (dashed line), the sudden transition disappears and the function 
becomes smoother. Correction ii), the impact of collisional ionizations (solid line), is significant 
at $z \gtrsim 6$ for $\Gamma_{\text{eff}}=10^{-24} s^{-1}$, but not for 
$\Gamma_{\text{eff}}=10^{-25} s^{-1}$. When the electron temperature reaches 
$\sim 10^4$ K, collisional ionizations become efficient, leading to a cooling of 
$T_{\text{e}}$ and an enhancement of $x_{\text{e}}$. As soon as all atoms are ionized, the 
electron temperature starts to increase again.  

\begin{figure}
\centering
	\includegraphics[width=0.6\textwidth]{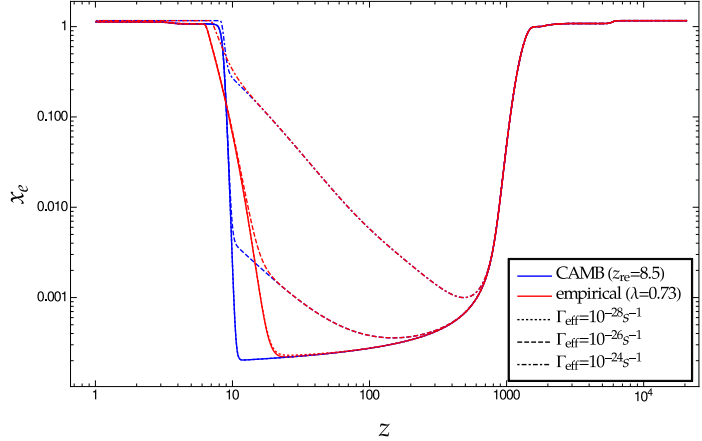}
\caption{
Effect of dark matter decay on the evolution of the free electron fraction. Solid lines are for 
stable DM, while broken lines show the effect of DM decay for 
$\Gamma_{\text{eff}}=10^{-28} \text{s}^{-1}$ (dotted), 
$\Gamma_{\text{eff}}=10^{-26} \text{s}^{-1}$ (dashed) 
and $\Gamma_{\text{eff}}=10^{-24} \text{s}^{-1}$ (dash-dotted).
The blue lines show the results for the CAMB parametrization \eqref{xe_standard} with $z_{\text{re}}=8.5$, 
the red lines for the empirical parametrization \eqref{xe_alternative} (red lines) with $\lambda=0.73$.}

\label{xe_plot}
\end{figure}

In figure \ref{xe_plot} we show the evolution of the free electron fraction for different values of 
$\Gamma_{\text{eff}}$ and for both scenarios of the astrophysical reionization. 
To obtain a smooth transition from CosmoRec to CAMB we add up the contributions from both codes as described in the appendix \ref{appendixA}. 
Note that with this procedure the interpretation of the evolution parameter $\lambda$ is different from the case without DM decay: 
For pure astrophysical reionization, $\lambda$ is a parameter that can directly be observed from astrophysical data. 
But when including DM decay the evolution of the free electron fraction is effectively described by astrophysical reionization and DM decay, the direct observation is hence a combination of $\lambda$ and $\Gamma_{\text{eff}}$. 
In this case $\lambda$ has to be interpreted as a parameter describing the astrophysical reionization process and not as a parameter describing direct observations.

\subsection{Impact of dark matter decay on CMB angular power spectra}
\label{Impact of dark matter decay on the CMB angular power spectra}

In figure \eqref{CMB_spectrum}, we show the TT, EE and TE angular power spectra for the CAMB parametrization \eqref{xe_standard} and the empirical parametrization \eqref{xe_alternative} of astrophysical reionization with and without DM decay, where we use Planck best fit values of the parameters \cite{Ade:2015xua}.

The impact of DM decay on the TT and high-$\ell$ EE angular power spectrum can roughly be understood in the same manner as the impact of astrophysical reionization described in section \ref{Impact of reionization on CMB angular power spectra}. Therefore, the differences at large $\ell$ that we see in this figure can be traced 
back to different values of the optical depth for the four considered models. The polarization spectrum at low $\ell$ in contrast does not only depend on the size of $\tau$, but also on the evolution of the free electron fraction. As expected, we observe that DM decay leads to an enhancement in the EE and TE spectra at intermediate and low $\ell$. We see that there are clear differences between 
the models for stable and unstable dark matter for all three spectra. Very extended reionization scenarios like DM decay lead to an enhancement of the polarization power at higher $\ell$ than rather sharp scenarios like astrophysical reionization. The importance of low-$\ell$ polarization data to constrain the DM decay rate $\Gamma_{\text{eff}}$ becomes evident. 

%%%%%%%%%%%%%%%%%%%%%%%%%%%%%%%%%%%%%%%%%%%%%%%%%%%%%%%%%%
\section{Constraints from CMB observations}
\label{Comparison with CMB data}

\subsection{Model fitting and comparison} 

The objective of this analysis is to study the impact of our assumptions on astrophysical reionization on 
the inference of cosmological parameters. In section \ref{Comparison of the two parametrizations 
of astrophysical reionization}, we therefore compare the two different parametrizations introduced in 
section \ref{Evolution of the free electron fraction}. We add DM decay as an additional source of 
reionization  and derive constraints on the DM decay rate using both parametrizations of 
astrophysical reionization in section \ref{Limits on dark matter decay rate}. This allows us to study the 
robustness of the constraints on the DM decay rate given the lack of information about 
astrophysical reionization. In section \ref{wdm}, we apply our analysis to a keV-mass sterile neutrino as a 
specific DM candidate and derive constraints on its mixing angle and mass.

\begin{table}
\begin{center}
\begin{tabular}{|c|c|c|c|c|}
\hline 
$\Omega_{\text{b}} h^2$ & $\Omega_{\text{c}} h^2$ & $\theta_{\text{s}}$ [deg] & $\ln(10^{10} A_{\text{s}})$ & $n_{\text{s}}$\\ 
\hline 
$0.02226$ & $0.1193$ & $1.04087$ & $2.0$ -- $4.0$ & $0.8$ -- $1.2$ \\
\hline 
\end{tabular} 
\end{center}
\caption{
Cosmological parameters of the flat $\Lambda$ cold dark matter models in our 
Markov chain Monte Carlo (MCMC) analysis. The dimensionless baryon and cold dark matter densities 
and the angular size of the acoustic sound horizon are fixed to their best-fit values from the 
Planck 2015 analysis. For the primordial scalar amplitude and the spectral tilt we indicate the 
range of the the flat prior distribution.}
\label{table_priors1}
\end{table}

We use the Bayesian approach to study the different reionization parametrizations with and without DM decay.
In order to find the posterior distributions of cosmological parameters, we use a modified version of the 
publicly available Markov Chain Monte Carlo (MCMC) parameter estimation code CosmoMC 
\cite{Lewis:2002ah}\footnote{Version July 2015.}, which makes use of the Boltzmann solver code CAMB \cite{Lewis:1999bs}\footnote{Version January 2015.}. 
CAMB adopts the recombination history from a library produced by the CosmoRec code 
\cite{Chluba:2005uz,Chluba:2010ca}\footnote{Version 2.0.3.}. The modules of reionization (CAMB) and recombination 
(CosmoRec) were respectively modified to include the empirical parametrization \eqref{xe_alternative} 
and the effect of DM decay, as described in sections \ref{Astrophysical reionization} and 
\ref{Dark matter decay and cosmic reionization}.

The CosmoMC code treats the reionization optical depth $\tau_{\text{reion}}$ as a free parameter and 
derives the redshift of reionization $z_{\text{re}}$ from it, using the parametrization \eqref{xe_standard}. 
Since we investigate the empirical parametrization \eqref{xe_alternative} within this work, treating 
$\tau _{\text{reio}}$ as a free parameter is impractical and we modified the code such that 
$z_{\text{re}}$ (or $\lambda$ in the empirical parametrization) is varied and we treat the optical 
depth as a derived parameter.

For the purpose of reionization analysis, it is not necessary to vary all six standard parameters of the flat 
$\Lambda$ cold dark matter ($\Lambda$CDM) model.
The dimensionless density of baryons ($\Omega_\text{b}h^2$), cold dark matter ($\Omega_\text{c}h^2$) and 
the angular diameter distance ($\theta_{\text{s}}$) have very little or no degeneracy with the degrees of 
freedom related to reionization \cite{Ade:2015xua}. 
The reason is that these parameters affect the CMB spectrum (position and amplitude of peaks and wells) in 
a scale dependent way. Reionization takes place well after recombination and affects all 
high multipole moments in the same way. Similar overall effects can be caused by changing 
the initial power spectrum, therefore the parameters $\ln(10^{10} A_{\text{s}})$ and $n_{\text{s}}$ show 
a significant degeneracy with the parameters related to reionization and are kept free in this analysis. 

The adopted values for the fixed parameters are kept to their best-fit values from the Planck 2015 
analysis \cite{Ade:2015xua}. Table \ref{table_priors1} shows those values as well as the prior 
ranges for the primordial power spectrum parameters of our analysis. Given our ignorance of the 
details of astrophysical reionization and the value of $\lambda$, we choose a flat prior in $\lambda$. 
Note that our choice of a flat prior for $z_{\rm re}$ means that the 
CAMB parametrization asks the question when does cosmic reionization happen, while a flat prior for $\lambda$ for the empirical 
parametrization asks how fast does it happen, since in the empirical parametrization we know already 
the redshift $z_{\rm p}$ when most of the reionization is completed.  
This has important consequences for the results of our analysis, as we discuss in section 
\ref{Comparison of the two parametrizations of astrophysical reionization}. 
Table \ref{table_priors2} describes the four MCMC runs with the respective adopted flat prior ranges 
of the parameters describing the different reionization parametrizations ($z_{\text{re}}$ and $\lambda$) and 
the decay of DM ($\Gamma_{\text{eff}}$). 

\begin{table}
\begin{center}
\begin{tabular}{|c|c|c|c|}
\hline 
CosmoMC run & $z_{\text{re}}$ & $ \lambda $ & $- \log_{10}$($\Gamma_{\text{eff}}\ \text{s}$) \\ 
\hline 
CAMB parametrization & $5.0$ -- $13.0$ & --- & --- \\
empirical parametrization & -- & $0.05$ -- $2.5$ & --- \\ 
CAMB parametrization + DM decay & $5.0$ -- $13.0$ & -- & $24$ -- $28$ \\ 
empirical parametrization + DM decay  & --- & $0.05$ -- $2.5$ & $24$ -- $28$  \\
\hline 
\end{tabular} 
\end{center}
\caption{Range of flat priors for the reionization and dark matter decay parameters used in the MCMC analysis. 
}
\label{table_priors2}
\end{table}

For the inference procedure, we use the Planck 2015 data \cite{Ade:2015xua} including three likelihoods: 
(i) the low-\textit{l} temperature and LFI polarization (bflike, $2 \leq \ell \leq 29$), 
(ii) the high-\textit{l} plik TTTEEE ($30 \leq \ell \leq 2508$) likelihood, and finally 
(iii) the lensing power spectrum reconstruction likelihood. Throughout this paper, this data set is called 
``Planck lowTEB \& TT, TE, EE \& lensing". The lensing reconstruction helps to fix the angular diameter 
distance and exempts the need for a low redshift measurement of standard markers such as baryonic acoustic 
oscillations or type Ia supernovae. This combination is among the data sets used in the reionization 
analysis contained in \cite{Ade:2015xua}, which should facilitate the comparison.  Planck 2016 intermediate data \cite{Aghanim:2016yuo,Adam:2016hgk} are not included, as they are not available for analysis yet.

We adopt the Gelamann-Rubin convergence criterion (variance of chain means divided by the mean of 
the chain variances) of $R-1<0.05$.

\begin{figure}
\centering
\includegraphics[width=1.\textwidth]{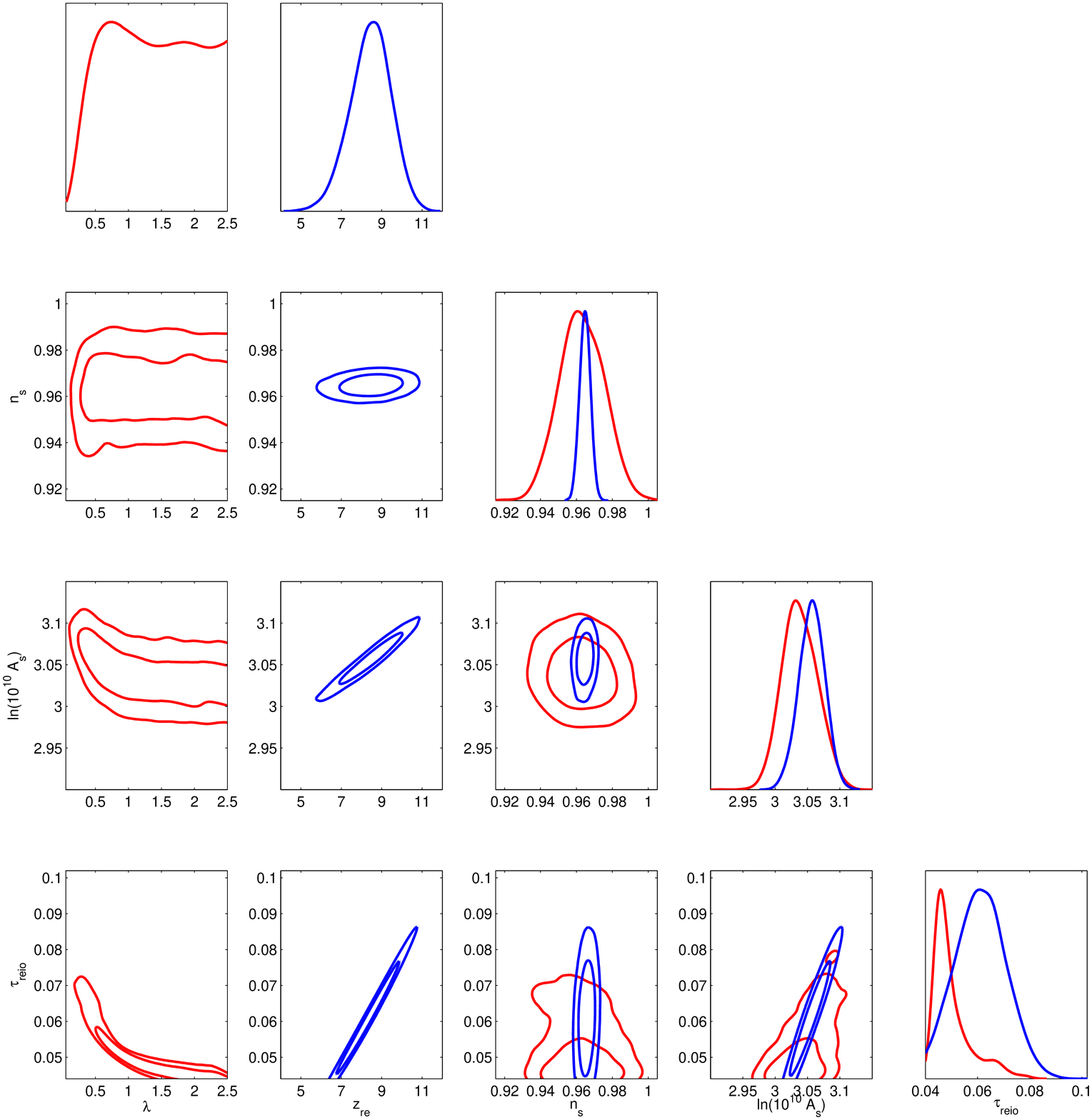}
\caption{Model constraints for stable dark matter. 2d marginalized 68\% and 95\% confidence 
contours and $1d$ marginalized posterior distributions are shown for the free parameters constrained 
using Planck 2015 data (TT, TE, EE, lowTEB, lensing). The empirical parametrization (red lines, $\lambda$) 
and the CAMB parametrization (blue lines, $z_{\text{re}}$) are compared for a flat $\Lambda$CDM model.}
\label{likelihoods_nodecay}	
\end{figure}

\subsection{Comparison of astrophysical reionization models}
\label{Comparison of the two parametrizations of astrophysical reionization}

In figure \ref{likelihoods_nodecay} we show the 68\% and 95\% confidence levels of 
$\ln (10^{10} A_{\text{s}})$, $n_{\text{s}}$ along with the empirical parametrization 
($\lambda$) as well as the CAMB parametrization ($z_{\text{re}}$). 
The corresponding mean averages and lower limits are listed in table \ref{table_results}.

We observe that models with $\lambda<0.37$ are excluded at 95\% confidence level which roughly 
translates into $\tau_{\text{reio}} \lesssim 0.07$. The CAMB parametrization in contrast shows 
\mbox{$z_{\text{re}} < 10.3$}, i.e.\ $\tau_{\text{reio}} \lesssim 0.08$. Very interestingly we furthermore 
find a significant enlargement of the posterior distribution of $n_{\text{s}}$ comparing the 
empirical parametrization to the CAMB parametrization in figure \ref{likelihoods_nodecay}. 

The reason for these findings --- tighter constraints on $\tau_{\text{reio}}$ and weaker constraints on 
$n_{\text{s}}$ --- lies within the different assumptions that we make when using the two different 
parametrizations of astrophysical reionization: 
The CAMB parametrization does not impose any prior knowledge about the redshift of reionization, but it 
makes assumptions about the process of reionization itself (i.e.~more or less instantaneous reionization). 
The empirical parametrization in contrast (with fixed $Q_{\text{p}}=0.99986$ and $z_{\text{p}}=6.1$) 
assumes some prior knowledge on the redshift of reionization, but allows for extended (small $\lambda$) 
as well as for sudden (large $\lambda$) reionization histories. Based on observations of quasars and 
star forming galaxies, the empirical parametrization assumes (almost) complete ionization for 
$z \leq 6.1$ and therefore has an intrinsic prior of $\tau_{\text{reion}} \gtrsim 0.04$. 

\begin{figure}
\centering
\includegraphics[width=\textwidth]{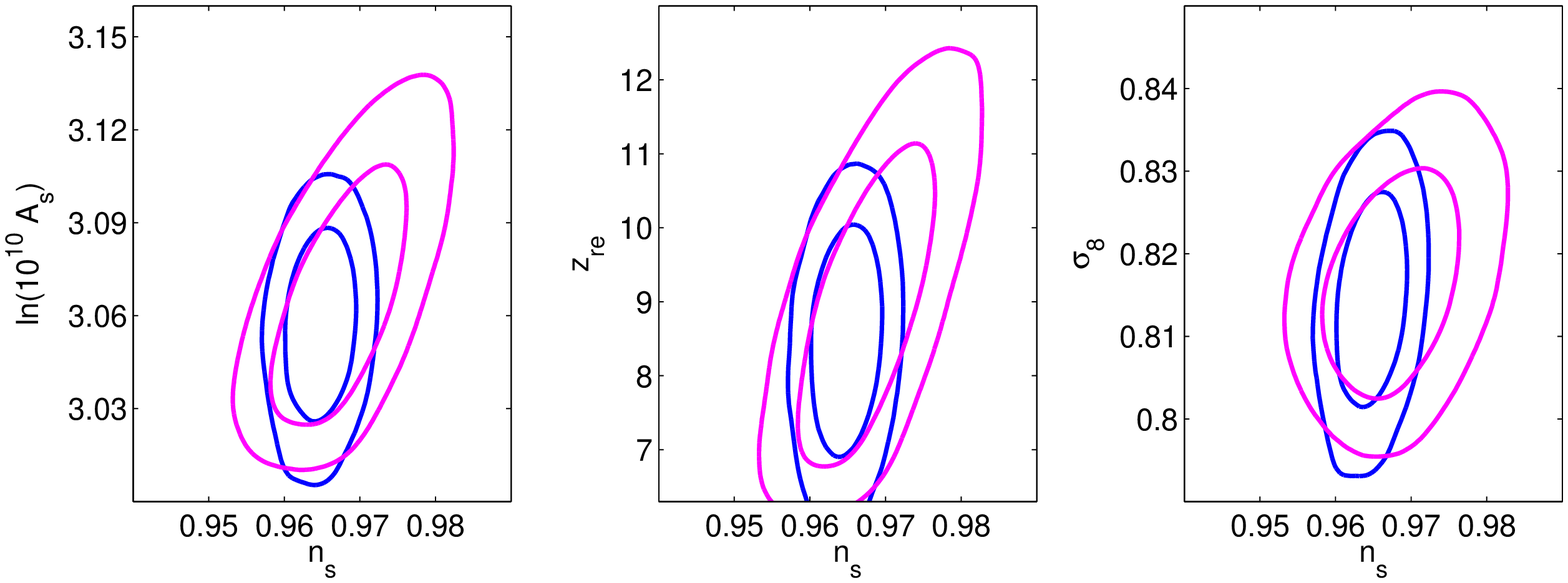}
\caption{2d marginalized constraints on cosmological parameters for stable dark matter. 
Blue lines: CAMB parametrization with the flat prior distribution for $z_{\text{re}}$ from table 
\ref{table_priors2}. Magenta lines: CAMB parametrization with a flat prior on the optical depth and  
$\tau_{\text{reio}}>0.04$.}
\label{likelihoods_alt}	
\end{figure}

The effect of implementing a prior of $\tau>0.04$ in the CAMB parametrization is 
shown in figure \ref{likelihoods_alt}.  Also in that case we observe a spreading 
of the posterior of $n_{\text{s}}$. However, this effect alone would not explain why high 
values of the optical depth are disfavored for the empirical parametrization.

The preference of low values of the optical depth is in fact implied by our choice of a flat prior on the 
evolution parameter $\lambda$. A flat prior in $z_{\text{re}}$ does in contrast lead to a relatively flat 
prior on the optical depth. We explain this in detail in appendix \ref{appendixB}.
From the perspective of CMB data analysis, with $\tau_{\text{reio}}$ as a principal component of the CMB, 
one may argue that it is desirable to use a flat prior on $\tau_{\text{reio}}$. 
However, from the perspective of reionization physics it appears more natural to us to assume a flat 
prior on $\lambda$, reflecting our ignorance of the evolution of the free electron fraction during reionization. 
The only thing we know about $\lambda$ is its order of magnitude (best fit 
$\lambda=0.73$ \cite{Douspis:2015nca}) and we chose its flat prior range such that it covers the 
same range of $\tau_{\text{reion}}$ as is covered by the flat prior range of $z_{\text{re}}$. 
A similar discussion but on the use of priors for $n_{\text{s}}$ and $A_{\text{s}}$ in inflationary models 
can be found in \cite{Valkenburg:2008cz}.

Finally, the combination of both effects, i.e.\ the prior of $\tau_{\text{reion}}>0.04$ (imposed by quasar 
and star forming galaxy observation) together with the preference of small values for 
$\tau_{\text{reion}}$ implied by a flat prior in $\lambda$, explains also the widening in the 
posterior of $n_{\text{s}}$. This is purely a normalization effect, as we show in more detail in 
appendix \ref{appendixB}. The interplay between $n_{\text{s}}$ and $\tau$ depending on the 
assumed reionization history is one of our main results.

\subsection{Limits on dark matter decay rate}
\label{Limits on dark matter decay rate}

\begin{figure}
\includegraphics[width=1.\textwidth]{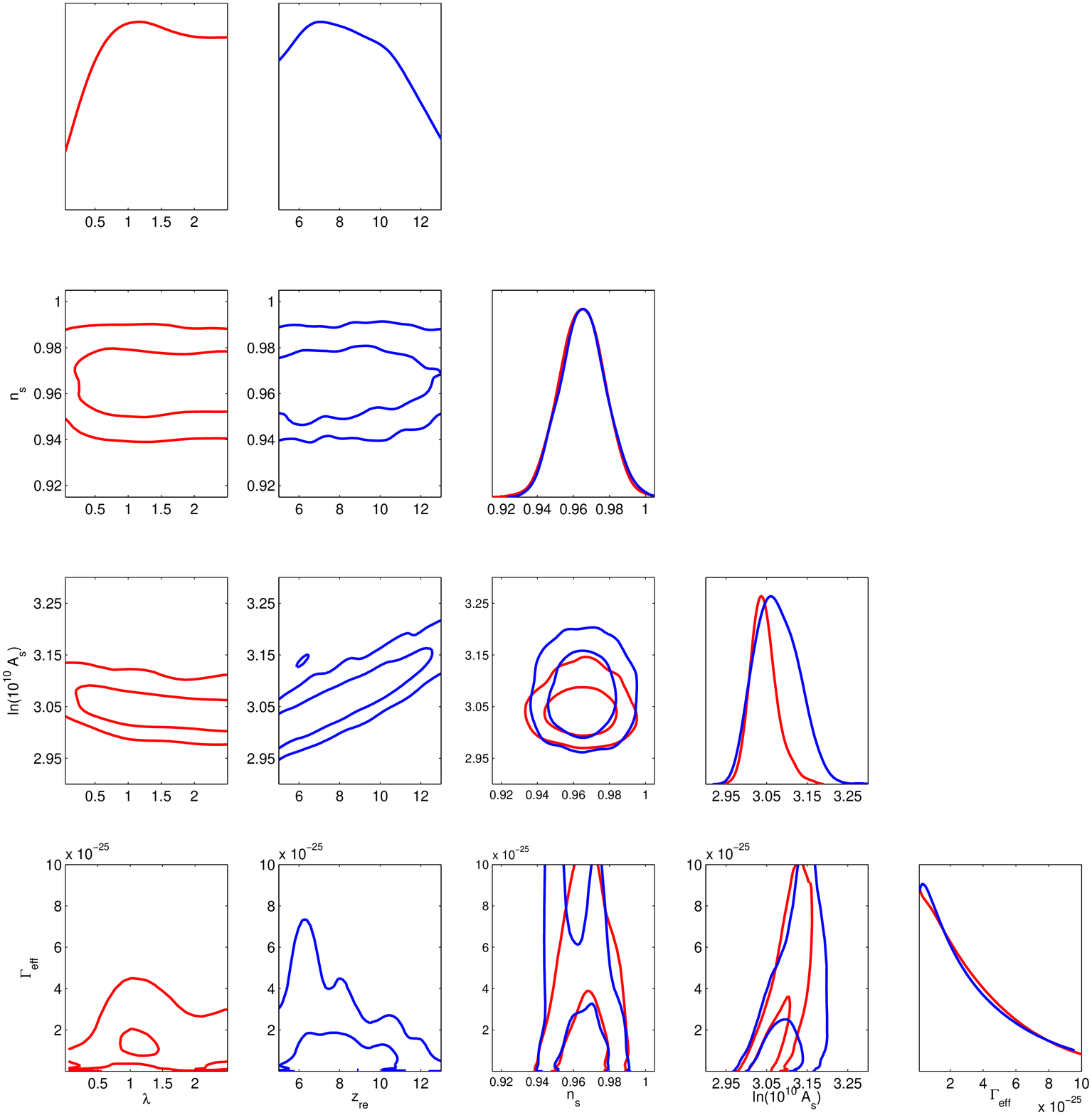}
\caption{Model constraints for decaying dark matter. 2d marginal 68\% and 95\% confidence contours 
and $1d$ marginalized posterior distributions are shown for the free parameters constrained using 
Planck 2015 data (TT, TE, EE, lowTEB, lensing). The empirical parametrization (red lines, $\lambda$) and 
the CAMB parametrization (blue lines, $z_{\text{re}}$) are compared when the dark matter decay rate 
$\Gamma_{\text{eff}}$ is added as an additional parameter.}
\label{likelihoods_decay}	
\end{figure}

\setlength{\tabcolsep}{3.3pt}
\renewcommand{\arraystretch}{1.1}

\begin{table}
\begin{center}
\begin{tabular}{|c|c|c|c|c|c|c|}\hline
CosmoMC run 	& $\ln(10^{10}\! A_{\text{s}})$ & $n_{\text{s}}$ & $z_{\text{re}}$
	& $ \lambda $ & $- \log_{10}$($\Gamma_{\text{eff}}\, {\rm s}$) & $\Gamma_{\text{eff}}$ [$s^{-1}$] \\ \hline
CAMB 			& $3.05^{+0.04}_{-0.03}$	& $0.965^{+0.006}_{-0.007}$ &
	$8.4^{+1.9}_{-2.0}$ 	& --- 	& --- 	& ---	\\ \hline
empirical  		& $3.04^{+0.05}_{-0.05}$	& $0.96^{+0.03}_{-0.02}$ & 
	--- & $>0.37$	 & --- & ---	\\ \hline
CAMB ~\& decay  	& $3.07^{+0.11}_{-0.09}$	& $0.97^{+0.02}_{-0.03}$ &
	${8.6^{+1.7}_{-3.1}}*$ & --- & $>24.59$ 	& $<2.6\!\times\!10^{-25}$	\\ \hline
empirical ~\& decay 	& $3.04^{+0.08}_{-0.06}$	& $0.96^{+0.03}_{-0.02}$ & 
	--- & $>1.7*$ & $>24.54$  & $<2.9\!\times\!10^{-25}$ \\ \hline
\end{tabular}
\end{center}
\caption{Constraints on cosmological parameters based on Planck 2015 data 
(TT, TE, EE, lowTEB, lensing), for the CAMB and empirical parametrizations for 
astrophysical reionization along with the constraints on the decay rate of dark matter.
We indicate the central value and the $95$\% confidence interval or the $95$\% lower or 
upper limits respectively. For the cases marked by an asterisk (*) the $68$\% interval or limit is 
quoted.} 
\label{table_results}
\end{table}

Let us now focus on the case of DM decay, see figure \ref{likelihoods_decay}. 
We find $\Gamma_{\text{eff}}< 2.6 \times 10^{-25} s^{-1}$ for the CAMB parametrization and 
$\Gamma_{\text{eff}}< 2.9 \times 10^{-25} s^{-1}$ for the empirical parametrization. 
Since these constraints agree within $\sim 10 \%$, we conclude that the constraints on DM 
properties show only a weak dependence on the chosen parametrization of reionization. 
This is reassuring, as it confirms that even though the details of astrophysical reionization are 
still widely unknown the constraints on DM properties are nevertheless reliable. 
The robustness of the bounds can be explained by the following considerations: Early reionization, like DM decay, leads to an enhancement of the TE and EE spectra in the intermediate $\ell$ range ($\ell \approx 10-60$). In contrast, late astrophysical reionization enhances the spectra only at lower $\ell$ (see figure \ref{CMB_spectrum}). For the same reason we also expect our limits on $\Gamma_{\text{eff}}$ to be weaker if the TE+EE data were excluded, see \cite{Ade:2015xua} for a discussion about the dependence of the constraints on the DM annihilation rate on the TE+EE data.

Our limits on $\Gamma_{\text{eff}}$ also rule out the scenario of pure DM reionization that is complete at $z_{\text{p}}=6.1$. This can be seen as a positive evidence for an astrophysical reionization process.

It is also interesting to note that when including DM decay there is a remarkable enlargement of the 
$n_{\text{s}}$-likelihood for the CAMB parametrization compared to the case without DM decay 
(figure \ref{likelihoods_nodecay}), whereas for the empirical parametrization it remains roughly 
the same. This can be explained by the fact that DM decay introduces extended 
reionization histories, a feature that cannot be mimicked by the 
CAMB parametrization alone but is to some extent already included in the empirical 
parametrization. 

Note that our constraints refer to the effective DM decay rate $\Gamma_{\text{eff}}$ \eqref{Gamma'}. 
If we assume that the decaying DM species makes up all DM of the universe, 
$(\rho_{\text{d}} /\rho_{\text{b}}) \approx 5.5$, the constraints on $\Gamma_{\text{eff}}$ are 
translated into $\Gamma \lesssim 5.3 \times 10^{-26} s^{-1}$ for the real DM decay rate. This constraint
assumes that all of the mass of the decaying particle goes into electromagnetic components and 
contributes to the ionization at time scales well below a Hubble time.  
 
We should also point out again that we work within the on-the-spot approach which assumes entirely 
efficient energy deposition, see section \ref{On-the-spot approximation}. 
Since this approximation in general holds only for high redshifts, the constraints presented in this 
section are overestimated and have to be treated with care. 
It is a common practice, e.g. \cite{Diamanti:2013bia,Ade:2015xua}, to go beyond the on-the-spot 
approximation by including a DM mass dependent constant instead of taking into account the full 
redshift dependence of $f_{\text{m}}(z,\Delta E_{\text{d}})$. 
This kind of effective treatment can also be applied to our constraints on $\Gamma_{\text{eff}}$, the 
prefactor that depends on the specific DM model can be calculated by codes like \cite{Slatyer:2015kla}. 

In the next section however, we consider the keV-mass sterile neutrino as a specific DM candidate, 
derive constraints on its decay rate and thereby go beyond the on-the-spot approximation.

\subsection{Constraints on sterile neutrino dark matter}
\label{sec:wdm}

An interesting candidate for decaying DM is the sterile neutrino with masses of the order of keV. 
Several production mechanisms for sterile neutrino DM have been proposed (see \cite{Adhikari:2016bei} for a review). 
From the constraints on the decay rate of dark matter, we derive model independent constraints on sterile neutrino parameters. 
In order to compare our bounds with existing limits (some of them model dependent), 
we illustrate these constraints in the context of the already ruled out non-resonant freeze-in production of sterile 
neutrino (Dodelson-Widrow model) \cite{Dodelson:1993je}. We also assume that sterile neutrinos are
the only form of DM.

According to the Dodelson-Widrow model, sterile neutrinos are produced via non-resonant chiral oscillations of the 
left-handed neutrinos, forming a warm\footnote{The notion of warm DM refers to the behaviour of the equation 
of state at the onset of structure formation, not to the question if they are thermal or non-thermal.} component of DM, 
which manifests itself as a strong suppression of structure below the free streaming length of sterile neutrinos. 
This suppression can be used to impose lower limits on the sterile neutrino mass by measuring structures at very small 
scales, using e.g.~surveys of Lyman-$\alpha$ forest or kinetic equilibrium of dwarf galaxies. 

Sterile neutrinos at this mass scale also have a radiative decay channel 
($\nu_s\rightarrow \gamma+\nu_\alpha$), emitting an active neutrino and a photon, each with an energy 
equal to half of the sterile neutrino mass. The produced X-ray photon can be directly measured or, given 
its absence, can be used to impose upper limits on the decay rate. The decay rate 
$\Gamma_{\nu_s \rightarrow \gamma\nu_\alpha}$ in turn is related to the sterile neutrino mass 
$m_s$ and mixing angle with active neutrinos $\theta$ in the following way
\begin{equation}
 \Gamma_{\nu_s \rightarrow \gamma\nu_\alpha} = \frac{9\alpha G_F^2}{256 \times 4\pi^4} 
 \sin(2\theta)^2 m_s^5 \ ,
\label{eq_wdm}
\end{equation}
where $\alpha$ is the fine structure constant and $G_F$ the Fermi constant. A complete and 
comprehensive review on keV sterile neutrino DM can be found in \cite{Adhikari:2016bei}.

Alternatively to direct observations, we can use the effect that the emitted X-rays would have on reionization. 
We derive constraints on the sterile neutrino decay rate $\Gamma_{\nu_s \rightarrow \gamma\nu_\alpha}$ 
and reinterpret them in terms of the mass and the mixing angle of the sterile neutrino. 
As shown in the previous section \ref{Comparison of the two parametrizations of astrophysical reionization}, the 
constraints on $\Gamma_{\text{eff}}$ are independent of the chosen parametrization of 
astrophysical reionization. Hence we consider in this part of our analysis the empirical parametrization only, 
since it gives the most conservative constraint on the decay rate. 

\begin{figure}
\centering
\includegraphics[width=0.9\textwidth]{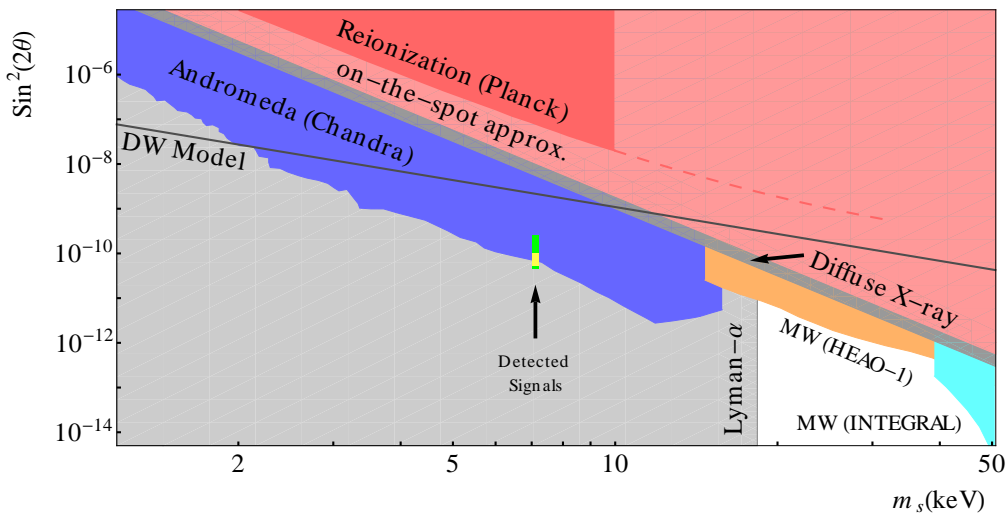}
\caption{Illustration of reionization limits on decaying dark matter. We plot the parameter space of 
freeze-in keV sterile neutrino dark matter. The light red area labelled ``on-the-spot approx." is excluded by the decay rate constraint 
$\Gamma_{\text{eff}}<2.9 \times 10^{-25}/$s obtained in section \ref{Comparison of the two parametrizations of 
astrophysical reionization}. The dark red area takes corrections to the on-the-spot approximation 
\cite{Ripamonti:2006gq} into account, valid for $2 \leq m_s/ \text{keV} \leq 10$. The dashed red line 
extrapolates the latter constraint to $m_{\text{s}}>10$ keV. Other regions of parameter space are 
excluded by the Tremaine-Gunn condition on the kinetic stability of dwarf galaxies ($m_\text{s} <  0.5$ keV is excluded) \cite{Tremaine:1979we},
the diffuse X-ray background \cite{Boyarsky:2005us}, the observed flux of X-rays from Andromeda \cite{Watson:2011dw} 
and the Milk Way \cite{Boyarsky:2006fg,Boyarsky:2007ge}, and a Lyman-$\alpha$ limit on the suppression of small scale 
structure for non-resonant production based on the bounds from \cite{Viel:2013apy} and the conversion formula from \cite{Bozek:2015bdo}. 
The latter limit is only valid for the Dodelson-Widrow (DW) non-resonant freeze-in model \cite{Dodelson:1993je}, 
which is depicted by the black line. For other models of sterile neutrino production, the Lyman-$\alpha$ limit 
depends not only on $m_\text{s}$ but also on $\theta$ \cite{Schneider:2016uqi}.  The claimed evidence for keV sterile 
neutrinos is shown in yellow \cite{0004-637X-789-1-13} and green \cite{Boyarsky:2014jta}.}
\label{wdm}	
\end{figure}

In order to model the effect of keV sterile neutrino decay on reionization in a realistic way, we have to go 
beyond the on-the-spot approximation which is only sufficient for deposited energies $<1$ keV 
\cite{Slatyer:2009yq}. 
The function $f_{\text{m}}(z,\Delta E_{\text{d}})$ that describes the energy deposition, see 
e.q.~\eqref{E_dep2}, can be evaluated numerically for each mass and redshift with the code 
of \cite{Slatyer:2015kla}, but for our purpose it is much more convenient to use the fitting 
formula derived in \cite{Ripamonti:2006gq},
\begin{equation}
f_\text{m}(z,\Delta E_{\text{d}})=\left[ 0.5+0.032 \left( \frac{m_s}{8 \text{keV}} \right)^{1.5} \right] \left[ \frac{z}{110 \left( \frac{m_s}{8 \text{keV}} \right)^{2.4}+z}\right]^{0.93}.
\label{fitting}
\end{equation}
This formula is valid for masses $2 \leq m_s/ \text{keV} \leq 10$. 
Including function \eqref{fitting} results in a reduction of the free electron fraction compared to the 
on-the-spot approximation,\footnote{$f_\text{m}(z,\Delta E_{\text{d}})=\frac{1}{2}$ since half of the 
energy is lost in form of neutrinos.} where the reduction is more pronounced at low redshifts and 
high masses $m_s$. 

We included eq. \eqref{fitting} into our implementation in CosmoRec and obtain mass dependent 
constraints on $\Gamma_{\text{eff}}$ which can be described by the fitting formula
\begin{equation}
 \Gamma_{\nu_s \rightarrow \gamma\nu_\alpha}(m_s) < 10^{-24}\left[1.29 + 2.11\times 10^{-2} \left( \frac{m_s}{\text{keV}} \right) + 1.48\times 10^{-2} \left( \frac{m_s}{\text{keV}} \right)^2  \right] \ \ \text{s}^{-1} \ ,
\end{equation}
which is valid for $2 \leq m_s/ \text{keV} \leq 10$. 

The corresponding constraints on the mixing angle $\theta$ and the 
mass $m_\text{s}$ are shown as the dark red area dubbed ``Reionization (Planck)" in figure \ref{wdm}, whereas 
the light red area represents the constraints using the on-the-spot approximation ($\Gamma_{\nu_s \rightarrow \gamma\nu_\alpha} <  5.3 \times 10^{-26}$s$^{-1}$). 
The red dashed line is a polynomial extrapolation (of 3rd order) of our results from $m_{\text{s}} \leq 10$ keV to higher masses in order to indicate the tendency of the constraint.
The constraints derived from reionization are weaker than the 
constraints from the diffuse X-ray background \cite{Boyarsky:2005us}. However, they seem to be competitive 
and especially promising given the perspectives on the sensitivity to observe the reionization history 
by future surveys \cite{Koopmans:2015sua}. 

The model dependent cases of non-resonant oscillation \cite{Dodelson:1993je},
resonant production \cite{Shi:1998km,Schneider:2016uqi} or decays of frozen-in scalars 
\cite{Merle:2013wta,Merle:2015oja} into sterile neutrinos would imply different abundances 
(lines in the $\theta-m_s$ plane) as well as different constraints from structure 
formation as the ones by the Lyman-$\alpha$ forest. A complete analysis that covers all 
possible models is beyond the scope of this work. The excluded Dodelson-Widrow non-resonant 
freeze-in model \cite{Dodelson:1993je} is plotted as a benchmark solely. 
Nevertheless, the competitive constraint obtained on the decay rate using the effect 
on reionization can be easily mapped to different models.

A complementary study of sterile neutrinos and the reionization history can be found in \cite{Rudakovskiy:2016ngi}, 
where the effect of free-streaming sterile neutrinos in the astrophysical reionization process itself was investigated.

%%%%%%%%%%%%%%%%%%%%%%%%%%%%%%%%%%%%%%%%%%%%%%%%%%%%%%%%%%
\section{Conclusions}
\label{Conclusions}

In this work, we studied the impact of DM decay on the CMB considering two different parametrizations for astrophysical reionization -- the conventional parametrization used by the CAMB code \eqref{xe_standard} (CAMB parametrization) and a recently proposed parametrization \cite{Douspis:2015nca} based on astrophysical observations \eqref{xe_alternative} (empirical parametrization). For equal values of the optical depth the empirical parametrization shows notable differences in the low-$\ell$ EE angular power spectrum of the CMB compared to the CAMB parametrization. Considering the decay of a DM species as an additional source of reionization, the CMB angular power spectra are furthermore sensitive to an effective DM decay rate $\Gamma_{\text{eff}}$ \eqref{Gamma'}. This effective decay rate includes not only the decay rate into electromagnetically or strongly interacting particles, but also factors characterizing the specific DM decay model. We modified the CosmoRec code to include the effect of DM decay and thereby had to take into account additional effects that are not yet included in the Recfast++ runmode of CosmoRec, namely a correction of the photon ionization coefficient and collisional ionizations. 

We derived constraints on cosmological parameters using the Planck (2015 release) data \cite{Ade:2015xua} including the low-\textit{l} temperature and polarization likelihood, the high-\textit{l} TT+TE+EE likelihood and the lensing power spectrum reconstruction likelihood. We find $\lambda > 0.37$ at 95\% confidence level for the evolution parameter $\lambda$ which characterizes the empirical parametrization.
We furthermore find that the empirical parametrization allows a much wider range of the spectral index than the CAMB parametrization, namely $n_{\text{s}}= 0.96^{+0.03}_{-0.02}$ (empirical) in contrast to $n_{\text{s}}=0.965^{+0.006}_{-0.007}$ (CAMB) at 95 \% confidence level. On the other hand, the reionization optical depth is tighter constrained in the case of the empirical parametrization, $\tau_{\text{reio}}=0.05_{-0.010}^{+0.017}$, than in the case of the CAMB parametrization, $\tau_{\text{reio}}=0.06_{-0.02}^{+0.02}$. 
This can be explained by the fact that our choice of a flat prior in the evolution parameter $\lambda$ implies a non-flat prior on the optical depth with a preference of low values of $\tau_{\text{reio}}$.
Given our lack of knowledge about the value of $\lambda$, a flat prior appears reasonable to us.
Furthermore the empirical parametrization has an intrinsic prior of $\tau_{\text{reio}} > 0.04$ assuming 
complete reionization for $z<6.1$ which is supported by observations of quasars \cite{Bouwens:2015vha}. 
We showed that applying a flat prior with $\tau_{\text{reio}} > 0.04$ also in the case of the CAMB 
parametrization results in a spreading of the likelihood distribution of $n_{\text{s}}$. The preference of low 
values of $\tau_{\text{reio}}$ together with the intrinsic prior on $\tau_{\text{reio}}$ in the empirical 
parametrization result in the remarkable 
broadening of the $n_{\text{s}}$-likelihood.  We conclude that prior knowledge of 
$\tau_{\text{reion}}$ is likely to weaken the strong constraints on $n_{\text{s}}$ that were 
reported in \cite{Ade:2015xua}. This can have important consequences for constraining inflationary models.

Turning to the case of DM decay, we find $\Gamma_{\text{eff}} < 2.6 \times 10^{-25} s^{-1}$ using the 
CAMB parametrization and $\Gamma_{\text{eff}} < 2.9 \times 10^{-25} s^{-1}$ using the 
empirical parametrization at 95 \% confidence level. With an agreement of $\sim$ 10\% we conclude 
that the constraints of $\Gamma_{\text{eff}}$ are independent of the chosen parametrization of 
astrophysical reionization. For the electromagnetic decay of a single component DM scenario this translates into $\Gamma <4.7 \times 10^{-26} s^{-1}$ and $\Gamma <5.3 \times 10^{-26} s^{-1}$. These constraints are obtained using the on-the-spot approximation 
(energy emitted by the DM decay immediately absorbed by the medium) and therefore overestimate the effect of DM decay on reionization, more realistic constraints are expected to be weaker. 
Notably, the likelihood of $n_{\text{s}}$ is also widened for the case of the CAMB parametrization 
when DM decay is included. 

As a specific application of our work, we considered the decay of a keV-mass sterile neutrino which has 
recently been claimed to be detected at $3.5$ keV \cite{0004-637X-789-1-13,Boyarsky:2014jta}. To 
obtain realistic constraints we extended our work beyond the on-the-spot approximation by including a 
redshift and mass dependent absorption fraction that takes into account the redshifting of the emitted 
photons \cite{Ripamonti:2006gq}. The constraints on the decay rate were reinterpreted in terms of the 
mass  $m_{\text{s}}$ and the mixing angle $\theta$ of the sterile neutrino. Our constraints are 
weaker but on a competitive level with those from the diffuse X-ray \cite{Boyarsky:2005us}.

The recent Planck analysis includes robust low-$\ell$ polarization data and 
resolves the reionization bump \cite{Aghanim:2016yuo,Adam:2016hgk}. The new data point to a slightly smaller value of the optical depth ($\tau_{\text{reio}}=0.58 \pm 0.012$, lollipop+TT) than the previous data ($\tau_{\text{reio}}=0.63 \pm 0.014$, TT,TE,EE+lowP+lensing) \cite{Ade:2015xua}. This is also consistent with our results for the empirical parametrization. Including the improved low-$\ell$ data will thus give rise to stronger constraints on the effective DM decay rate. 

%%%%%%%%%%%%%%%%%%%%%%%%%%%%%%%%%%%%%%%%%%%
\section{Acknowledgements}
We are thankful to Jens Chluba for his help with the implementation in CosmoRec, Sebastien Clesse for 
helpful hints about CosmoMC, Antony Lewis for useful suggestions about sharp edges, Alexander Merle and Aurel Schneider
for fruitful discussions about sterile neutrinos. Numerical calculations have been performed using JARA-HPC resources at the RWTH
Aachen Compute Cluster. 
DB acknowledges financial support from CAPES-CSF, grant 8768-13-7, and from the 
``Bielefeld Young Researchers' Fund". IMO furthermore acknowledges financial support from 
 Studienstiftung des Deutschen Volkes. We acknowledge support from RTG 1620 ``Models of Gravity" funded 
by DFG.

%%%%%%%%%%%%%%%%%%%%%%%%%%%%%%%%%%%%%%%%%%%

\appendix
\section{Smoothness of visibility function}
\label{appendixA}
Within the line-of-sight approach \cite{Seljak:1996is}, the CMB anisotropies depend not only on the visibility function $g=\dot{\tau} e^{-\tau}$, but also on its first and second derivatives with respect to cosmic time, $\dot{g}$ and $\ddot{g}$. Therefore, it is important to ensure that the free electron fraction is sufficiently smooth everywhere. There are two transitions in which discontinuities can appear in our model of the free electron fraction described in sections \ref{Astrophysical reionization} and \ref{Dark matter decay and cosmic reionization}:

\begin{itemize}
\item[i)] At the transition redshift $z_{\text{p}}=6.1$ of the empirical parametrization \eqref{xe_alternative}. To avoid the sharp edge in $x_{\text{e}}$, we modify the parametrization \eqref{xe_alternative} according to \footnote{A. Lewis, private communication.}
\begin{equation}
x_{\text{e}} (z) = 1.08 \times
\begin{cases} \frac{1-Q_{\text{p}}}{(1+z_{\text{p}})^3-1} \left( (1+z_{\text{p}})^3-(1+z)^3 \right) +Q_{\text{p}} \hspace{1cm} & \text{for } z < z_{\text{p}}\\
Q_{\text{p}}  e^{-\lambda \frac{(z-z_{\text{p}})^3}{(z-z_{\text{p}})^2+0.2}} & \text{for } z \geq z_{\text{p}}. \end{cases} 
\label{smooth_Q}
\end{equation}
In figure \ref{xe_comparison} we show the original parametrization (black lines) and its smoothed version \eqref{smooth_Q} (red lines). The smoothing procedure alters the function slightly at all redshifts, but the basic asymptotic behaviour remains the same. Being precise, equation \eqref{smooth_Q} still does not have a continuous derivative in a mathematical sense, but it is sufficiently smooth to remove any unphysical spikes as shown in figure \ref{visibility}. We show the visibility function and its first two derivatives, where the solid lines represent the smoothed function and the dotted ones the sharp function. The sharp edge of the original parametrization \eqref{Q} leads to spikes that are most dramatic in the second derivative of the visibility function and can lead to unphysical bias.
\begin{figure}[h!]
\centering
	\begin{subfigure}[b]{0.6\textwidth}
		\includegraphics[width=\textwidth]{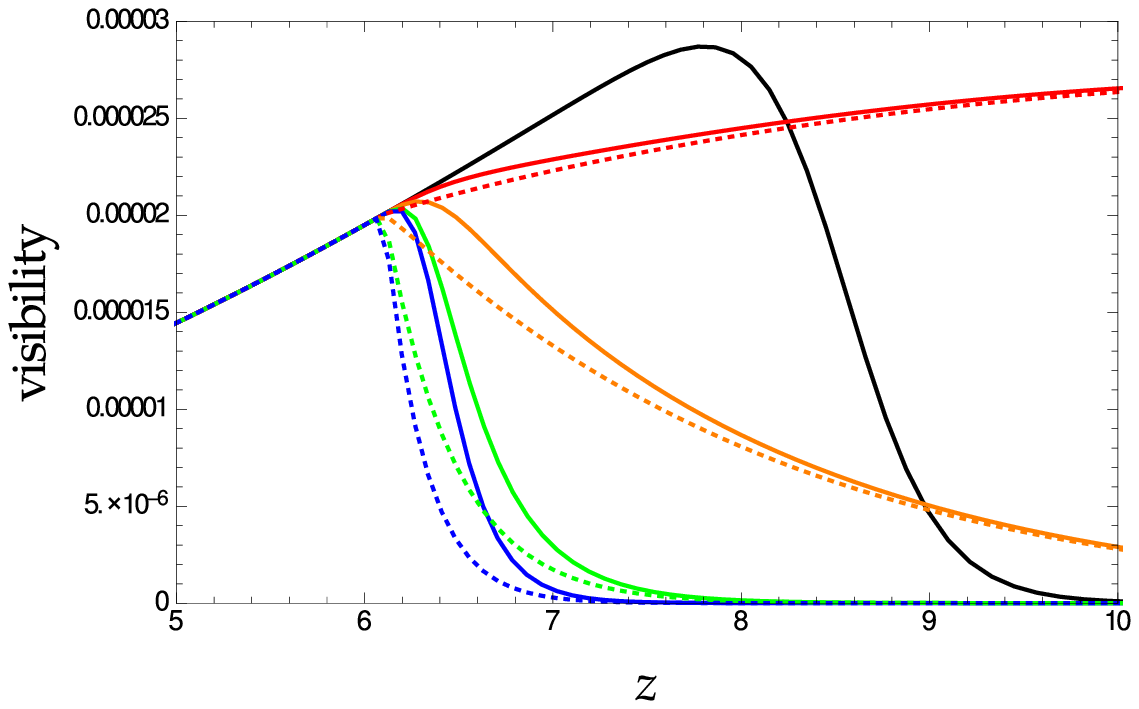}
	\end{subfigure}
	\begin{subfigure}[b]{0.6\textwidth}
		\includegraphics[width=\textwidth]{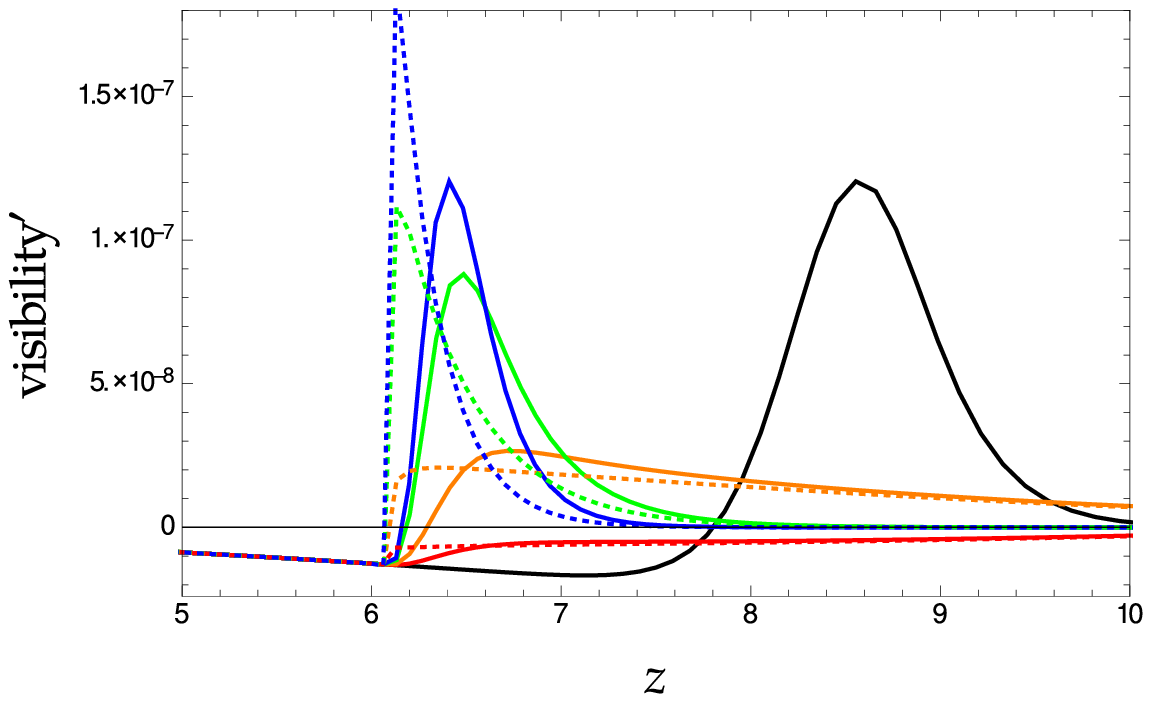}
	\end{subfigure}	
	\begin{subfigure}[b]{0.6\textwidth}
		\includegraphics[width=\textwidth]{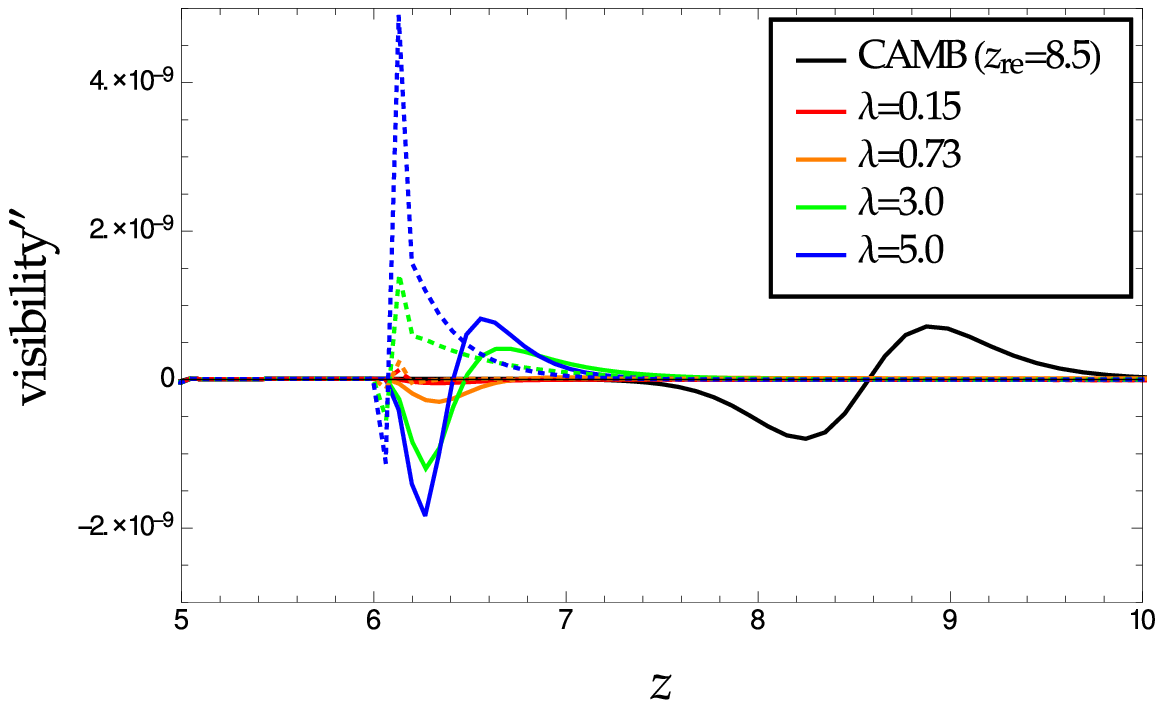}
	\end{subfigure}
\caption{Visibility function (top), its first (middle) and second derivative (bottom) of the empirical 
parametrization \eqref{xe_alternative} for different values of $\lambda$, where the dotted curves 
represent the sharp function \eqref{Q} and the solid curves its smoothed counterpart \eqref{smooth_Q}. 
The black curves display the CAMB parametrization for reference.}
\label{visibility}
\end{figure}		

\item[ii)] At the transition from recombination to astrophysical reionization (CosmoRec to CAMB modules). For astrophysical reionization histories and using the CAMB parametrization \eqref{xe_standard}, the code ensures a smooth transition from recombination to astrophysical reionization. The code uses the output of CosmoRec until a redshift of $z_{\text{s}}=z_{\text{re}}+8 \Delta_z$, from where the CAMB code takes on the control of the evolution of the free electron fraction in order to initiate the astrophysical reionization. At this redshift the remaining free electrons from recombination have reached an almost constant level of $\sim10^{-4}$ and the hyperbolic tangent \eqref{xe_standard} from the astrophysical reionization is still small and flat enough such that the transition is guaranteed to be smooth.

However, for the case of DM decay we cannot use the same procedure, because in this way DM decay is not included at low redshifts. The same also applies to studies of DM annihilation.  This is not only undesirable from a physical point of view, but it can also lead to a sharp edge at the transition redshift $z_{\text{s}}$ which in turn leads to unphysical spikes in the derivatives of the visibility function. In order to avoid this problem we start to add up the recombination (CosmoRec) contribution and the reionization (CAMB) contribution at sufficiently high redshifts. This procedure results in a smooth transition, as can be seen in figure \ref{xe_plot}. But we also have to make sure that the free electron function reaches the correct asymptotic value of $1.16$ (assuming double helium ionization), since a simple addition of both contributions would result in higher values. The naive way of setting a minimum condition, i.e. $x_{\text{e}}=\text{min}(1.16, \,x_{\text{e}}|_{\text{CAMB}}+x_{\text{e}}|_{\text{CosmoRec}})$, would again lead to a sharp feature. We solve this problem by the following smoothing function,
\begin{equation}
x_{\text{e}}= 1.16 \times \left[ \tanh \left( \left[ \frac{x_{\text{e}}|_{\text{CAMB}}+x_{\text{e}}|_{\text{CosmoRec}}}{1.16} \right]^{n} \right) \right]^{1/n},
\end{equation} 
where we choose $n=8$ and which behaves as a simple addition as long as the argument of the hyperbolic tangent is $\leq 1$ and then smoothly merges the two contributions to the asymptotic value of $1.16$.
\end{itemize}

\section{Optical depth prior}
\label{appendixB}
In this appendix we discuss the different priors on the optical depth imposed by using different parametrizations of reionization, i.e. \eqref{xe_standard} and 
\eqref{xe_alternative}, and their implications for the inference of the spectral index $n_{\text{s}}$. Using a flat prior in $\lambda$ for the empirical parametrization 
results in a different prior on $\tau_{\text{reion}}$ when compared to a flat prior in $z_{\text{re}}$ for the CAMB parametrization. 
The relation between the prior on the optical depth $P(\tau_{\text{reio}})$ and the flat prior $\tilde{P}$ in $z_{\text{re}}$ and $\lambda$ is respectively given by the following 
relations,

\begin{equation}
P_{\text{CAMB}}(\tau_{\text{reio}}) = \tilde{P}(z_{\text{re}})  \left| \frac{\mathrm{d} z_{\text{re}}}{\mathrm{d}  \tau_{\text{reio}}} \right| \hspace{1cm} \text{and} \hspace{1cm}
P_{\text{emp.}}(\tau_{\text{reio}}) = \tilde{P}(\lambda)  \left| \frac{\mathrm{d} \lambda}{\mathrm{d}  \tau_{\text{reio}}} \right|.
\label{prior_relation}
\end{equation}

In order to find $P_{\text{CAMB}}(\tau_{\text{reio}})$ and $P_{\text{emp.}}(\tau_{\text{reio}})$ we have to derive $z_{\text{re}}(\tau_{\text{reio}})$ and 
$\lambda(\tau_{\text{reio}})$, which we can obtain by evaluating the integral in \eqref{optical_depth} at $z=z_{\text{\text{rec}}}$, i.e., up to the redshift of recombination.
For simplicity we hereby neglect the ionization of HeII at redshift $\sim$ 3.5 and furthermore the contribution of the cosmological constant $\Lambda$ to the Hubble parameter.

\begin{figure}[t]
\centering
	\includegraphics[width=0.55\textwidth]{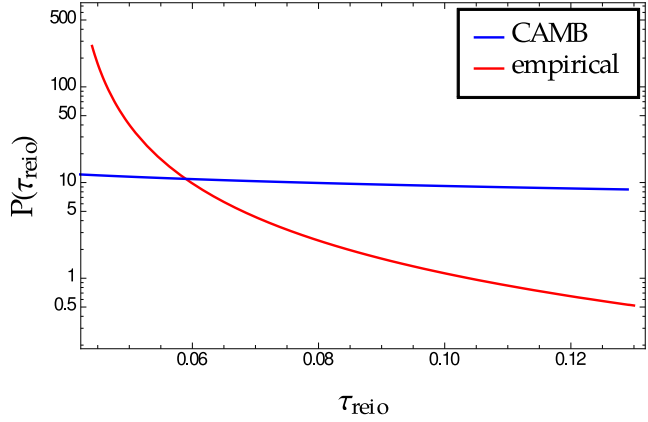}
\caption{Prior distribution of $\tau_{\text{reio}}$ implied by a flat prior on 
$z_{\text{re}}$ for the CAMB parametrization \eqref{xe_standard} (blue line) 
and a flat prior on $\lambda$ for the empirical parametrization \eqref{xe_alternative} (red line) 
using eq. \eqref{prior_relation}.}
\label{Priors}
\end{figure}	

For the CAMB parametrization it is convenient to approximate eq. \eqref{xe_standard} by a stepfunction, which is sufficient to study the leading order dependence 
of $\tau_{\text{reio}}$ on $z_{\text{re}}$. We hence find

\begin{equation}
\begin{aligned}
\tau_{\text{reio}} \big \vert_{\text{CAMB}} &= c \sigma_T \int_0^{z_{\text{rec}}} \frac{n_{\text{e}}(z')}{(1+z') H(z')}  \, \mathrm{d}z' \approx \frac{1.08 c 
\sigma_T n_{\text{H,0}}}{H_0 \sqrt{\Omega_{\text{m}}}} \int_0^{z_{\text{re}}} \sqrt{1+z'} \, \mathrm{d} z' \\
& = \frac{2}{3} \alpha  \left[ (1+z_{\text{re}})^{3/2} -1 \right],
\end{aligned}
\end{equation}
where we defined $\alpha= \frac{1.08 c \sigma_T n_{\text{H,0}}}{H_0 \sqrt{\Omega_{\text{m}}}}$. This immediately implies 

\begin{equation}
\frac{\mathrm{d}z_{\text{re}}}{\mathrm{d}\tau_{\text{reio}}} \propto \left( \tau_{\text{reio}} + \frac{2}{3} \alpha \right)^{-1/3}. 
\label{dzre_dtau}
\end{equation}

For the empirical parametrization we find instead

\begin{equation}
\begin{aligned}
\tau_{\text{reio}} \big \vert_{\text{emp.}} & \approx \alpha \left [ \int_{0}^{z_{\text{p}}} \left( \frac{1-Q_{\text{p}}}{(1+z_{\text{p}})^3-1} 
\left( (1+z_{\text{p}})^3-(1+z')^3 \right) + \, Q_{\text{p}} \right) \sqrt{1+z'} \, \mathrm{d} z'  \right.\\ 
& \left. \hspace{1cm} + Q_{\text{p}}\int_{z_{\text{p}}}^{\infty} e^{-\lambda(z'-z_{\text{p}})} \sqrt{1+z'} \, \mathrm{d} z' \right] \\
& = \beta + \frac{\alpha Q_{\text{p}} e^{\lambda(1+z_{\text{p}})}}{\lambda^{3/2}} \Gamma \left( 3/2,\lambda(1+z_{\text{p}}) \right),
\end{aligned}
\label{tau_emp}
\end{equation}
where $\beta$ is a constant that is simply defined by the first integral in \eqref{tau_emp} and $\Gamma$ is the incomplete Gamma function. 
Note that for the second integrand of \eqref{tau_emp} we have approximated \mbox{$z_{\text{rec}} \rightarrow \infty$}. 
It is possible to differentiate \eqref{tau_emp} with respect to $\lambda$ analytically, but rewriting the result in terms of $\tau_{\text{reio}}$ has to be done numerically. 
In figure \ref{Priors} we show the priors on $\tau_{\text{reio}}$ using a flat prior in $z_{\text{re}}$ (eq. \eqref{dzre_dtau}) and a flat prior in $\lambda$ (eq. \eqref{tau_emp}),
taking into account the corresponding normalization.

\begin{figure}[t]
\centering
	\includegraphics[width=0.55\textwidth]{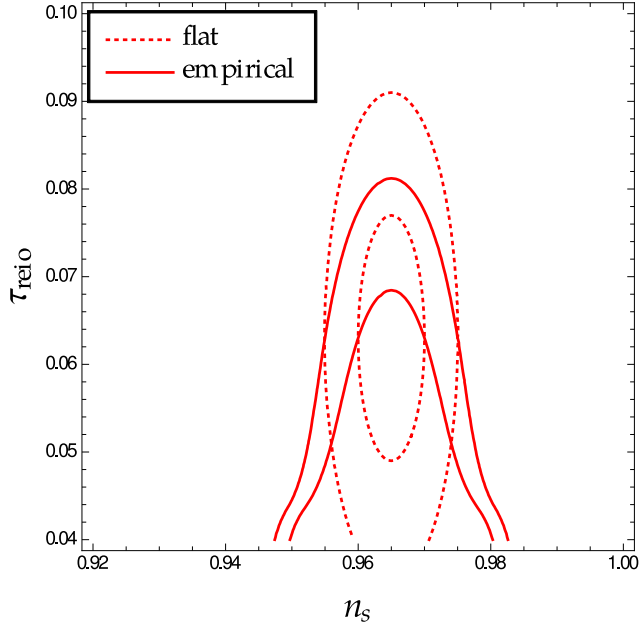}
\caption{68\% and 95 \% confidence contours of a binormal distribution in the 
$\tau_{\text{reio}}$-$n_{\text{s}}$ plane (dashed lines). After multiplying with the prior 
distribution $P_{\text{emp.}}(\tau_{\text{reio}})$ the resulting normalized posterior distribution 
(solid lines) broadens in $n_{\text{s}}$ direction.}
\label{Normalization}
\end{figure}	

The empirical parametrization implicitly restricts the optical depth to values $\tau_{\text{reio}} \gtrsim 0.044$, since it assumes 
(almost) complete ionization for $z\leq z_{\text{p}}=6.1$. On the other hand, as we can see in figure \ref{Priors}, a flat prior in $\lambda$ results in a preference for 
small values of $\tau_{\text{reio}}$. The CAMB parametrization instead has a relatively flat prior in $\tau_{\text{reio}}$ when assuming a flat prior in $z_{\text{reio}}$. 
This explains why we find much tighter constraints on $\tau_{\text{reio}}$ when using the empirical parametrization compared to the CAMB parametrization, 
see figure \ref{likelihoods_nodecay} in section \ref{Comparison of the two parametrizations of astrophysical reionization}. 

The tight constraints on $\tau_{\text{reio}}$ in turn result in much weaker constraints on $n_{\text{s}}$ and also $A_{\text{s}}$, as shown in figure \ref{likelihoods_nodecay}. 
To understand this effect let us for simplicity focus on the $\tau_{\text{reion}} - n_{\text{s}}$ parameter space and neglect the dependence on $A_{\text{s}}$ for the moment. 
As we show schematically in figure \ref{Normalization} the widening of the posterior distribution is simply caused by normalization. 
We show in blue a mock likelihood distribution for $\tau_{\text{reion}}$ and $n_{\text{s}}$, assuming a Gaussian normal distribution in both parameters. 
When multiplying the Gaussian likelihood distribution by the prior distribution $P_{\text{emp.}}(\tau_{\text{reio}})$ (figure \ref{Priors}) and renormalizing accordingly, 
the posterior distribution gets squeezed into $n_{\text{s}}$ direction. 

The same argument holds also in the $\tau_{\text{reio}}$-$A_{\text{s}}$ plane and we observe the same widening in $A_{\text{s}}$, 
see figure \ref{likelihoods_nodecay}. However, the difference between the CAMB \eqref{xe_standard} and the empirical parametrization \eqref{xe_alternative} in the \textit{marginalized} 
posterior distribution is much less pronounced for $A_{\text{s}}$ than for $n_{\text{s}}$. This can be explained by the degeneracy of $\tau_{\text{reio}}$ with $A_{\text{s}}$, 
whereas $\tau_{\text{reio}}$ and $n_{\text{s}}$ are basically uncorrelated.

\bibliographystyle{utcaps}
\bibliography{LiteratureDMDecay}

%%%%%%%%%%%%%%%%%%%%%%%%%%%%%%%%%%%%%%%%%%%%%%%%%%%%%%%%%%%%%%%%%%%%%%%%%%%%%%%%%%%%%

\end{document}